# Quantum Amplitude Estimation for Ligand and Pose Index Retrieval

Pei-Kun Yang

Independent Researcher, Taiwan

Email: peikun@isu.edu.tw, peikun6416@gmail.com

**Abstract**

Structure-based virtual screening spans a combinatorial configuration space, motivating a quantum screening framework that targets threshold-qualified configuration indices without reconstructing a complete interaction-score table. Quantum amplitude estimation based on quantum phase estimation (QAE/QPE) supplies finite-resolution phase information to a reversible comparator that writes its decision to FLAG, while optional Grover-style quantum selection (QSEL) amplifies FLAG-selected outcomes and selector measurements identify the ligand, receptor state, rotation, and translation. In an ideal logical benchmark using AutoDock grid interactions as a proof-of-concept model, we evaluated initial FLAG acceptance probability, classification precision, shots required for 95% FLAG detection, and logical $Q$-operator calls across eight receptor systems and nested libraries of up to 1,024 ligand occupancies. Higher QAE/QPE resolution generally improved classification precision at additional evaluation cost, whereas suitably chosen QSEL iteration counts reduced required shots without improving classification precision, revealing selection and resource trade-offs for extending coherent index retrieval to more physically detailed molecular evaluation models.

## Introduction

Structure-based virtual screening (SBVS) seeks to identify ligands capable of binding favorably to a target protein and to determine plausible binding configurations. In physical terms, the corresponding thermodynamic criterion is the standard binding free energy $\Delta G_{\text{bind}}$, defined by the equilibrium between bound and unbound states. Statistical thermodynamics provides a rigorous framework for this quantity through partition functions of the interacting system, including the protein, ligand, solvent, and ions. The energies entering these statistical weights ultimately originate from the underlying molecular interactions and, at the electronic level, quantum mechanics. Binding free energy is therefore a theoretically defined quantity rather than an empirically invented target. Accurate practical evaluation nevertheless requires the energetic and entropic contributions of vast collections of microscopic states [1-3].

For ligand prioritization, the ideal approach from a physical standpoint would be to compare sufficiently accurate binding free energies for the relevant ligands, accounting for their accessible molecular states. Such a comparison would rank compounds according to the intended thermodynamic property, subject to the modeled solution conditions and chemical states. Most routine SBVS calculations cannot perform this task. Instead, they estimate tractable interaction energies, empirical docking scores, or learned affinity surrogates, which can support screening but are not interchangeable with rigorous $\Delta G_{\text{bind}}$ values. Accordingly, selecting the ligand with the best computed docking score does not establish that it has the most favorable true binding free energy. The distinction reflects the approximations needed to make large-scale screening computationally feasible [4-7].

A major obstacle is computational scale. Drug-like chemical space is enormous, and databases of commercially accessible compounds already contain many billions of candidates [8]. For each ligand, binding depends on more than molecular identity. The receptor can adopt multiple conformations, a flexible ligand can adopt many internal conformations, and the ligand can occupy different translations and orientations within the binding region. These degrees of freedom are coupled through the interaction energy, yet the number of possible combinations grows multiplicatively as more states are included. Even if a numerical energy could be evaluated rapidly for one configuration, extending the calculation across a sufficiently complete receptor–ligand–pose ensemble would require a vastly greater number of evaluations. Although computational advances can accelerate the evaluation of individual configurations, the number of possible receptor–ligand configurations continues to grow combinatorially as additional molecular states and pose variables are considered [9-12].

Practical SBVS therefore relies on computationally tractable approximations. Simplified interaction models and empirical scoring functions enable efficient evaluation of molecular configurations, while machine-learning-based scoring provides an additional approach for assessing candidate poses [3-5, 13, 14]. These approximations are useful, but their computational efficiency is often obtained by limiting physical detail or the configurations examined.

Structural flexibility introduces a further source of combinatorial growth. A receptor may undergo side-chain or backbone rearrangements, while ligand torsions and solvent-mediated interactions affect the accessible bound configurations. Conventional docking often samples only a subset of these possibilities. Flexible-side-chain docking, ensemble docking, and induced-fit docking extend that coverage but remain constrained by available computing resources. Excluding receptor or ligand states before evaluation can omit favorable combinations, since a poor score for one mismatched configuration does not establish that the ligand itself is unfavorable [9-12, 15-18].

The same limitation applies to ligand poses. Because translational, rotational, and torsional variables generate a large search space, docking programs generally rely on heuristic sampling and local optimization rather than exhaustive evaluation [13, 19, 20]. Methods that identify promising placements before detailed evaluation, including grid sampling guided by quadratic unconstrained binary optimization and geometric triplet matching, reduce the number of poses to be examined [21]. Such advances improve the exploration of selected regions but do not remove the need to evaluate many molecular configurations. Benchmark measures of docking accuracy, enrichment, and ranking assess performance under specific datasets and evaluation protocols, but do not establish exhaustive coverage of all possible molecular configurations [5, 22].

The resulting computational challenge has two distinct components. First, physically detailed interactions may need to be evaluated over an immense set of molecular configurations. Second, the useful screening output must be extracted from the evaluations. Conventional approaches address the first challenge through approximate scores and restricted sampling. A quantum approach invites a different representation of the configuration space, while also requiring a strategy for retrieving useful screening results rather than a prohibitively large numerical score table.

A coherent quantum computational framework could, in principle, apply a molecular energy-evaluation operation across many indexed configurations within a single quantum state. Suppose the evaluator requires $d$ working qubits and an additional $s$ selector qubits to label $2^s$ receptor, ligand, and pose configurations. If the required controlled evaluator and state preparation are available, the same logical operation can act coherently on the indexed superposition. This offers a long-term route for combining large molecular configuration spaces with increasingly accurate physical models. Quantum parallelism, however, does not make all numerical outputs simultaneously accessible. Reconstructing a score to comparable precision for every configuration generally requires a total measurement budget that increases with the number of indexed configurations. This readout problem motivates a different screening output.

Our earlier research addressed complementary parts of quantum SBVS. One study explored compact quantum machine learning for binding-affinity prediction. Subsequent studies established grid-based receptor–ligand interaction energies as quantum-compatible inner products and extended the representation to indexed receptor, ligand, rotation, and translation configurations [23-25]. Those grid-based studies recovered numerical interaction energies from Hadamard-test measurement statistics. The present work advances the same SBVS objective by changing both the coherent overlap encoding and the required output. The interaction is encoded as a designated amplitude available to quantum amplitude estimation based on quantum phase estimation (QAE/QPE) [26]. A reversible quantum comparator (QCOMP) marks configurations that satisfy a prescribed threshold, while optional Grover-style quantum selection (QSEL) amplifies the accepted subspace [27, 28]. Measurement of FLAG and the selector registers then yields a candidate ligand and pose index rather than a reconstructed score for every configuration. We assess FLAG acceptance, finite-resolution classification precision, the shots required for candidate detection, and logical $Q$-operator calls. An eight-receptor benchmark with nested ligand libraries evaluates these properties using selected AutoDock4 grid-interaction terms as a controlled proof-of-concept model.

## Methods

**Notation and screening objective.** Each selector index specifies a receptor state, ligand occupancy, rotation, and translation. Let $N_{\mathrm{OCC}}$ denote the number of ligand occupancies. The indexed search size is $N_{\mathrm{search}} = 32{,}768\ N_{\mathrm{OCC}}$, and $M$ denotes the number of threshold-qualified configurations. The normalized grid-interaction amplitude is $A_0$, and $A_{0,\mathrm{thr}}$ is its screening threshold. The qualifying event $G$ is $A_0 < A_{0,\mathrm{thr}}$. Under uniform selector preparation, $q = M/N_{\mathrm{search}}$ is the reference qualifying fraction in the numerical benchmark. QCOMP writes its finite-resolution threshold decision to FLAG. The initial FLAG acceptance probability $p = \Pr(\mathrm{FLAG} = 1 \mid r = 0)$ is the probability of observing FLAG = 1 in a single execution before QSEL. Under uniform selector preparation, $p$ is also the expected fraction of selector samples accepted by QCOMP at the chosen screening threshold and evaluation resolution. With sufficiently accurate threshold classification, $p$ approaches the proportion of configurations whose scores satisfy that threshold. Classification precision $\eta = \Pr(G \mid \mathrm{FLAG} = 1)$ is the fraction of FLAG-selected outcomes that satisfy the threshold in the benchmark. This measure quantifies statistical candidate reliability under the tested conditions rather than certifying any individual retrieved index. The screening threshold $A_{0,\mathrm{thr}}$ and evaluation-register size $m$ are independently specifiable. For numerical analysis, we use a base setting $m_0$ and a resolution multiplier $k$ to organize their variation, as detailed below. With $r$ nonnegative QSEL iterations, $p_r = \Pr(\mathrm{FLAG} = 1 \mid r)$ is the FLAG acceptance probability after QSEL. Ideal QSEL changes this probability without changing the screening threshold or $\eta$. $S_{95}(r)$ is the minimum number of independent shots required to observe FLAG = 1 at least once with 95% probability, and $C_{95}(r)$ is the corresponding logical $Q$-operator call count. Each call denotes one application of the internal Grover operator used in QAE/QPE. For complete score readout, $S_{\text{score}}$ is the mean sampling budget for $N_{\text{search}}$ numerical scores, each requiring $s(\varepsilon)$ index-associated observations at precision $\varepsilon$. After FLAG = 1 is observed, measuring the selector registers returns a candidate index. Figure 1 identifies the circuit registers. Supplementary Notes S1 and S2 detail the energy encoding and general probability model.

**Molecular input preparation and receptor selection.** Molecular inputs were obtained from PDBbind protein–ligand complexes [29, 30]. We selected a tractable proof-of-concept test set whose structures could be processed into the Protein Data Bank with partial charges and atom types (PDBQT) format. Systems that could not complete the preparation workflow were excluded. Ligand structures were obtained in structure data file (SDF) format when available and in MOL2 format otherwise. They were then converted to rigid PDBQT using Meeko [31].

Receptor selection was constrained by the downstream grid and quantum encoding. Pose evaluation used a 32 × 32 × 32 lattice with a spacing of 0.375 Å and included ligand rotation and translation. To retain sufficient spatial margin for these pose operations, the native ligand was restricted to a minimum enclosing sphere diameter of no more than 7.5 Å, corresponding to 20 grid intervals. A three-qubit channel index represented eight interaction channels comprising one electrostatic channel and seven AutoDock atom type channels. The seven AutoDock atom types that collectively covered the largest number of eligible PDBbind ligands were A, C, HD, N, NA, OA, and S. Native ligands used for receptor selection were therefore required to contain only atom types from this set.

Among eligible systems, we favored receptors with larger native ligands within the 7.5 Å diameter limit to increase the potential receptor–ligand interaction contributions. Both a ligand-bound complex and the corresponding protein-only system were required for receptor state generation. Systems for which CHARMM-GUI preparation could not be completed were excluded. Because each receptor required two molecular dynamics (MD) simulations and molecular preparation was computationally demanding, the production set was limited to eight receptor systems, namely 1ecq, 1h0r, 2aa9, 2q9y, 2ybk, 2ybs, 4gui, and 6o4x.

**MD and receptor state generation.** For each selected system, a ligand-bound complex and the corresponding protein-only system were prepared independently with CHARMM-GUI [32, 33], solvated and ionized, and simulated with OpenMM [34]. The ligand-bound simulation was intended to retain a receptor conformation with favorable native receptor–ligand interactions, whereas the independently simulated protein-only system was used to obtain a receptor conformation relaxed in the absence of the bound ligand, which could differ in its complementarity to the native ligand. Protein interactions were described with the CHARMM36m force field [35], and ligand parameters in the complex simulations were taken from the CHARMM General Force Field [36]. Each system underwent 5,000 steps of energy minimization and was then equilibrated for 125 ps at 303.15 K using a 1 fs time step, Langevin dynamics with a friction coefficient of 1 $ps^{-1}$, constraints on bonds involving hydrogen atoms, and positional restraints of 400 and 40 kJ $mol^{-1}$ $nm^{-2}$ on protein backbone and side chain atoms, respectively. Electrostatics were treated with particle mesh Ewald summation, and no pressure coupling was applied during equilibration. Unrestrained production dynamics were then performed for 1 ns at 303.15 K and 1 bar using a 2 fs time step and an isotropic Monte Carlo barostat. Van der Waals (vdW) interactions used force switching from 0.8 to 1.0 nm for the ligand-bound complex and from 1.0 to 1.2 nm for the protein-only system. After production, the ligand-bound complex underwent an additional 5,000 steps of energy minimization, whereas the protein-only receptor was taken from the production endpoint. These short MD simulations were used to generate two contrasting receptor states for subsequent grid analysis, not to establish equilibrium conformational ensembles.

**Structural processing after MD.** Water molecules and common ions were removed from the structures after MD. The receptor and ligand were extracted from the additionally minimized ligand-bound complex, and the protein-only receptor was extracted from its production endpoint. The receptor pair and native ligand were translated together so that the ligand bounding box center coincided with the origin. A common half-grid shift of −0.1875 Å was then applied along each Cartesian axis. The protein-only receptor was then superimposed on the receptor derived from the complex by Kabsch alignment [37, 38]. The two aligned receptor structures were converted to PDBQT with AutoDockTools/MGLTools [39]. For the ligand, bond connectivity was retained from the corresponding SDF template while the atomic coordinates were replaced by those extracted from the processed complex. Atom counts, element order, and bond lengths were checked before conversion to rigid PDBQT using Meeko [31]. The processed native ligands were verified to remain within grid indices −12 to 11 along $x$, $y$, and $z$ for subsequent grid construction and pose operations.

**Candidate library construction.** We constructed nested ligand libraries using distractors selected by molecular size to examine index-first screening with sparse qualifying configurations. For each receptor, its corresponding native ligand was always retained as the reference occupancy. A common pool of distractor ligands was assembled from additional PDBbind entries. Distractors had to be convertible to rigid PDBQT and to contain only the seven selected AutoDock atom types A, C, HD, N, NA, OA, and S. Unlike the native ligands used in receptor selection, distractors were not restricted to a 7.5 Å minimum enclosing sphere diameter. Eligible distractors were ranked by decreasing diameter, and the first 1,023 were retained. The larger ligands were prioritized because placement in unrelated receptor environments was expected to increase steric overlap and unfavorable vdW interactions. Molecular size determined distractor selection, whereas the screening threshold was applied to $A_0$.

Each distractor was centered independently at its bounding box center and shifted by −0.1875 Å along each Cartesian axis before generating occupancy grids. For each receptor system, the ligand occupancy count $N_{\mathrm{OCC}}$ took the values 1, 2, 4, 8, 16, 32, 64, 128, 256, 512, and 1024. Selector 0 always represented the native ligand, and the remaining positions were filled by successively appending distractors in the predetermined size order. Both the qualifying count and $q$ could vary with library size.

**Receptor maps and ligand occupancy grids.** Receptor-dependent interaction maps were generated separately for the two aligned receptor states with AutoGrid4 using the AutoDock4 parameter set [39]. Grid calculations were centered at the origin with 0.375 Å spacing, 33 × 33 × 33 grid points, a smoothing parameter of 0.5 Å, and a dielectric value of −0.1465. The electrostatic map and seven ligand atom-type affinity maps were retained for scoring. Under the standard combined-map convention, these affinity maps include vdW, applicable hydrogen-bond, and charge-independent desolvation contributions. The separate AutoGrid4 charge-dependent desolvation map (d.map) and non-grid terms such as the torsional penalty were excluded from the benchmark inner product. Thus, the numerical energy is a selected grid score rather than the complete AutoDock4 score. The downstream encoded representation used 32 × 32 × 32 spatial arrays. The encoded arrays retained indices 0–31 along each axis of the original AutoGrid maps, omitting index 32 on each axis. Before the inner-product calculations, cropped receptor-map values were clipped to the interval [−1000, 1000] and stored to six decimal places. Ligand partial charges were projected onto the electrostatic channel and ligand atom occupancies onto the corresponding atom type channels by trilinear interpolation on the same lattice. Atoms outside the retained grid were ignored, and interpolation weights assigned to grid points outside the retained array were discarded. The interpolated ligand charge and atom-type occupancy grids were also stored to six decimal places. Subsequent normalization and interaction calculations used these stored values in double precision. The order of eight channels was e, A, C, HD, N, NA, OA, and S for every receptor and ligand, with absent channels filled with zeros. The resulting physical vector contained $8 \times 32^3 = 262{,}144$ values, corresponding to an 18-qubit physical data register.

**Coherent interaction energy encoding and auxiliary normalization.** The grid-interaction energy $E$ was represented by the inner product of receptor-map and ligand charge/occupancy vectors. One auxiliary qubit extended the physical vectors to a fixed 19-qubit encoded data space. Within each receptor system, the two receptor states shared the normalization factor $L_{\mathrm{U}}$, and the native ligand and all 1,023 distractors shared $L_{\mathrm{N}}$ across the nested libraries. A common energy offset $E^{+}$ for that receptor system shifted the indexed interactions into the nonnegative range. Orthogonal auxiliary components equalized the vector norms without changing $E$. The resulting amplitude was $A_0 = (E + E^{+})/(L_{\mathrm{U}}\, L_{\mathrm{N}})$. Supplementary Note S1 provides the generalized scoring-channel representation, normalization bounds, and auxiliary construction.

**Pose encoding and search space definition.** The rotation (ROT) and translation (TRANS) blocks in Figure 1 implement the quantum-compatible spatial transformations developed in our previous studies [24, 25]. The ROT block applies index-controlled permutations of Cartesian grid coordinates, while the TRANS block performs index-controlled cyclic shifts along the spatial axes. These operations transform the physical ligand occupancy grid while leaving the auxiliary components unchanged. Five rotation selector qubits encoded 32 selector states corresponding to 24 unique cubic orientations. Nine translation selector qubits encoded 512 cyclic translations, with three bits assigned to each Cartesian axis and integer shifts from −4 to +3 grid cells. The two receptor map states, together with $N_{\mathrm{OCC}}$ ligand occupancies, 32 rotation selector states, and 512 translations, define an indexed selector space of $N_{\mathrm{search}} = 2 \times N_{\mathrm{OCC}} \times 32 \times 512 = 32{,}768 \times N_{\mathrm{OCC}}$, ranging from 32,768 configurations at $N_{\mathrm{OCC}} = 1$ to 33,554,432 configurations at $N_{\mathrm{OCC}} = 1024$. The occupancy selector therefore required 0 to 10 qubits (Supplementary Note S3, Table S1). Together, the selector registers provide a complete label for each encoded configuration. Measuring these registers returns the occupancy index, receptor state, rotation, and translation needed to identify the ligand and reconstruct its encoded pose.

**Quantum circuit architecture.** Figure 1 summarizes the logical circuit. Hadamard gates prepared uniform superpositions over the receptor map, occupancy, rotation, and translation selectors. SELECT-OCC prepared the selected normalized ligand occupancy, the ROT and TRANS blocks applied the selected pose, and SELECT-MAP combined the posed occupancy with the selected receptor map state so that the designated amplitude $A_0$ encoded the normalized, configuration-dependent inner product between the receptor map and ligand occupancy. QAE/QPE then mapped the associated phase information to an evaluation register. QCOMP wrote the screening decision to FLAG, and optional QSEL amplified the subspace marked by FLAG. FLAG and selector measurements provide the acceptance decision and configuration index. The numerical interaction score is not read out.

The QAE/QPE block represents controlled powers of an internal Grover operator, with each configuration selector label preserved during phase estimation. QSEL uses a distinct Grover operator defined by the complete coherent preparation through QCOMP and its inverse, with FLAG = 1 as the marked subspace. Supplementary Note S2 details the unitary constructions of both operators. Figure 1 depicts their logical roles rather than a gate-level decomposition.

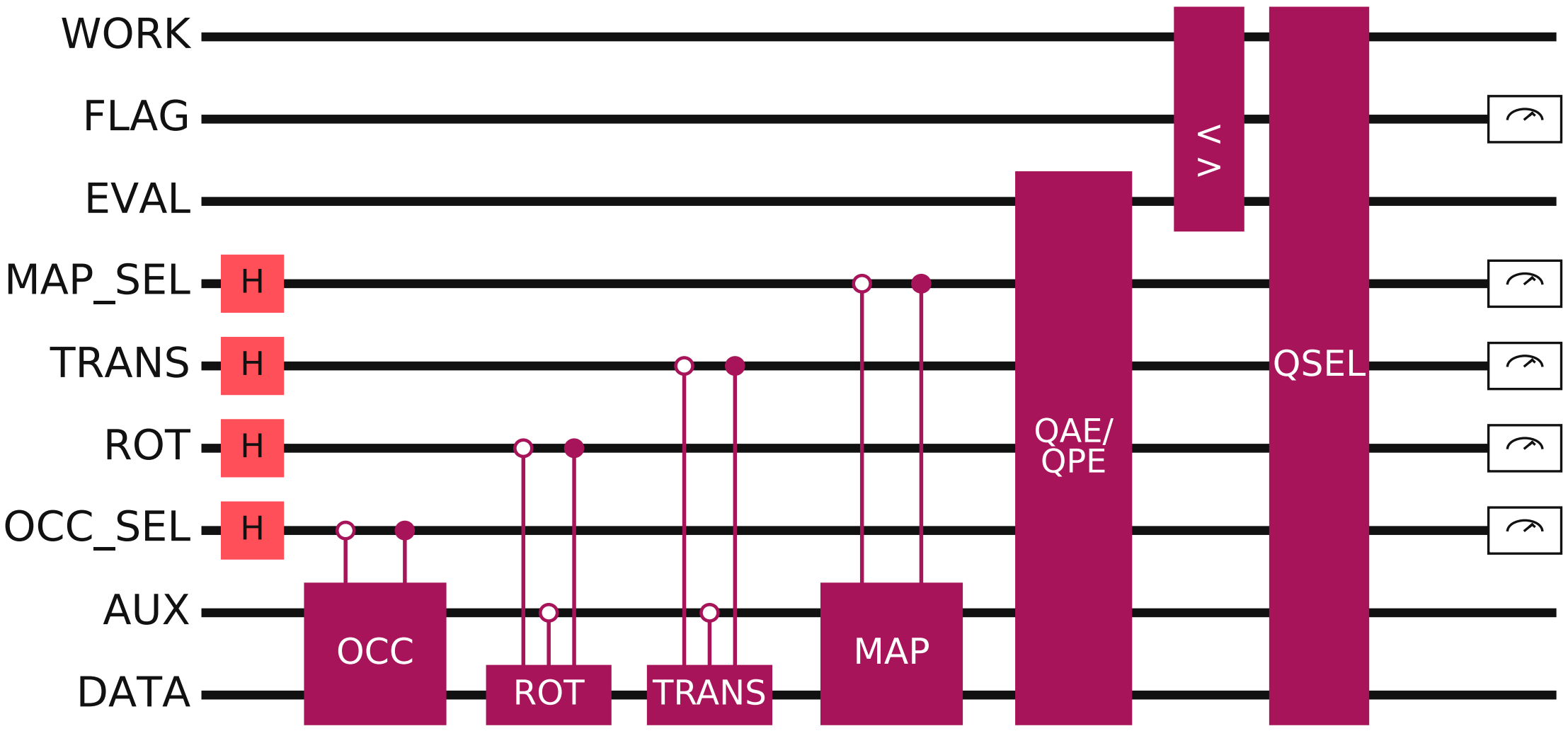

**Figure 1.** Selector registers prepared by Hadamard gates index receptor maps, ligand occupancies, rotations, and translations. The selected occupancy is transformed by pose operations and combined with the selected receptor map to encode the interaction amplitude. QAE/QPE maps the phase information associated with $A_0$ to the evaluation register, QCOMP writes the screening decision to FLAG, and optional QSEL amplifies the subspace selected by FLAG. Measurement of FLAG records acceptance, and measurement of the selector registers identifies the ligand, receptor state, rotation, and translation. The data and auxiliary wires are shown separately for clarity, although together they form the fixed 19-qubit encoded data space.

**Threshold qualification and finite-precision QAE/QPE classification.** QAE/QPE yields phase estimates associated with the two eigenphase branches corresponding to $\varphi$ and $1 - \varphi$, where $\varphi = \sin^{-1} A_0/\pi$ [26, 28]. QCOMP accepts the low- and high-phase regions corresponding to the screening threshold and writes its decision to FLAG using a reusable work register [40]. Finite evaluation resolution can produce false-positive and false-negative decisions. Increasing $k$ refines the phase estimate at fixed $A_{0,\mathrm{thr}}$.

**Readout comparison assumptions.** For complete score readout, $S_{\mathrm{score}} \approx N_{\mathrm{search}} \times s(\varepsilon)$ is the mean budget when each index needs $s(\varepsilon)$ associated observations at precision $\varepsilon$. The index-first task instead stops when FLAG = 1 is observed and the selector registers have been measured to retrieve a candidate ligand and pose index.

**QSEL amplification, FLAG-detection shots, and $Q$-operator costs.** Under ideal QSEL, the FLAG acceptance probability after $r$ iterations is $p_r = \sin^2\left[(2r+1)\sin^{-1}\sqrt{p}\right]$

[27, 28]. Ideal amplification preserves $\eta$ within that subspace. The corresponding 95% shot requirement is $S_{95}(r) = \lceil \ln(0.05)/\ln(1 - p_r) \rceil$ when $0 < p_r < 1$. One shot suffices for $p_r = 1$, while no finite count suffices for $p_r = 0$. Each QAE/QPE evaluation block uses $2^m - 1$ logical $Q$-operator calls. One initial block and two blocks per QSEL iteration yield $C_{95}(r) = S_{95}(r)(2r + 1)(2^m - 1)$. Each application of $Q$ or its inverse counts as one call.

**Logical resources and numerical parameter sweep.** The circuit includes the fixed 19-qubit encoded data register, one receptor-state selector qubit, up to 10 occupancy selector qubits, five rotation selector qubits, nine translation selector qubits, an $m$-qubit evaluation register, one FLAG qubit, and reusable QCOMP workspace. We used eleven nested libraries with $N_{\mathrm{OCC}}$ = 1, 2, 4, 8, 16, 32, 64, 128, 256, 512, and 1024. The parameter sweep used $m_0$ = 10–22 and $k$ = 1, 2, 4, and 8, setting $A_{0,\mathrm{thr}} = \sin(\pi/2^{m_0})$ and $m = m_0 + \log_2(k)$. This parameterization simplifies the QCOMP integer comparison. At fixed $m_0$, varying $k$ changes the evaluation resolution without changing the threshold. The corresponding evaluation-register settings are listed in Supplementary Table S2. $A_{0,\mathrm{thr}}$ ranged from approximately $3.07 \times 10^{-3}$ at $m_0 = 10$ to $7.49 \times 10^{-7}$ at $m_0 = 22$. The sampled $r$ values included 0 and extended beyond the first FLAG-amplification peak. The maximum sampled iteration count varied among resolution multipliers.

**Numerical implementation and model scope.** Pose-dependent interactions were evaluated in double precision by cyclic correlation using fast Fourier transform (FFT) methods. PyTorch was used for FFT calculations and NumPy for analytical probability calculations [41, 42]. The finite-precision QAE/QPE response, including both eigenphase branches, was evaluated analytically. We calculated $p$, $\eta$, $p_r$, $S_{95}(r)$, and $C_{95}(r)$ in an ideal noiseless logical model. At fixed $r$, eight-receptor medians and interquartile ranges (IQRs) were computed from receptor-specific results. For reported minimum costs, each receptor's $C_{95}(r)$ was minimized over its sampled nonnegative $r$ values, including zero, before taking the eight-receptor median and IQR. The 95% criterion in Figures 4 and 5 concerns at least one FLAG = 1 outcome. Candidate reliability is assessed separately by $\eta$. The reported $C_{95}(r)$ counts logical $Q$-operator calls rather than elementary quantum gates or hardware execution time.

The large-library benchmark used FFT-derived normalized interaction amplitudes, analytical finite-register QAE/QPE and QCOMP acceptance probabilities, and ideal QSEL amplification to evaluate the predicted FLAG probabilities and costs. Separate Qiskit validations examined an integrated 23-qubit prototype and a full-resolution 29-qubit ROT/TRANS subcircuit. Independent non-FFT calculations also checked the full-scale numerical overlaps (Supplementary Note S4).

## Results

**Measurement requirements.** Table 1 compares the outputs and shot requirements of two readout tasks. Complete score readout requires a numerical interaction score for every indexed configuration. For $N_{\text{search}}$ indices sampled uniformly, with a mean of $s(\varepsilon)$ index-associated observations per score at precision $\varepsilon$, the mean shot budget is $S_{\text{score}} \approx N_{\text{search}}\ s(\varepsilon)$. This budget increases with the size of the indexed library. Index-first screening instead returns a FLAG-selected ligand and pose index without reconstructing all scores. If the initial FLAG acceptance probability is $p$, the probability of obtaining at least one FLAG = 1 in $S$ independent shots is $1 - (1 - p)^S$. For $p \ll 1$, the 95% shot requirement is $S_{95}(0) \approx \ln(20)/p$. Ideal QSEL amplification can increase FLAG acceptance to $p_r$ and reduce $S_{95}(\text{r})$, potentially to constant order near an amplification peak. These requirements concern different outputs and do not constitute a comparison of total implementation time. Supplementary Note S2 provides the derivations and underlying assumptions.

Across the eleven nested libraries, $N_{\text{search}}$ ranges from 32,768 to 33,554,432 configurations as $N_{\text{OCC}}$ increases from 1 to 1024. At a fixed per-score observation budget, the mean complete-readout requirement therefore increases by a factor of 1024 across this range. The FLAG-selected retrieval criterion instead specifies the number of shots required to obtain an accepted index, without imposing reconstruction precision on every score.

**Table 1.** Measurement requirements for complete score readout and FLAG-selected candidate retrieval.

| **Output task** | **Shot requirement** | **Condition** |
|---|---|---|
| Complete score readout | $S_{\text{score}} \approx N_{\text{search}} \times s(\varepsilon)$ | Fixed per-score precision $\varepsilon$ |
| FLAG-selected candidate retrieval ($r$ = 0) | $S_{95}(0) \approx \ln(20)/p$ | $p \ll 1$ |
| FLAG-selected candidate retrieval with QSEL | $S_{95}(r) = O(1)$ | Ideal QSEL amplification, $r = O(\sqrt{1/p})$, with $r$ chosen near the first amplification peak |

Note. $S_{\text{score}}$ is a mean budget at per-score precision $\varepsilon$. The $S_{95}(0)$ approximation applies when $p \ll 1$. The QSEL row refers to an ideal amplification setting, with the iteration count chosen near the first FLAG-amplification peak. Both candidate retrieval rows target at least one FLAG = 1 outcome with 95% probability. Supplementary Note S2 gives the FLAG-detection probabilities and logical-cost derivations.

**Reduced-scale gate-level validation.** We tested the logical architecture in Figure 1 using a reduced-scale gate-level implementation. A 23-qubit implementation in Qiskit included selector-controlled occupancy and receptor-map preparation, rotation, translation, QAE/QPE, reversible QCOMP accepting both eigenphase regions, and QSEL. Statevector simulations agreed with independent analytical calculations for the component operations and the FLAG and index outputs. Supplementary Note S4 presents the circuit schematic (Figure S1) and the validation comparisons (Table S3). The eight-receptor benchmark, including its shot requirements and logical $Q$-operator call counts, was computed classically using the ideal circuit probability model.

**Full-resolution ROT/TRANS gate-level validation.** A separate 29-qubit Qiskit gate-level simulation implemented the 32 × 32 × 32 spatial transformations with 15 coordinate qubits, five rotation-selector qubits, and nine translation-selector qubits. Across four initial coordinates, all 32 × 512 selector combinations were tested (65,536 cases), and every output coordinate matched an independently computed reference. A separate coherent interference test involving 16 branches also passed, with a return probability of 1.000000000000 and leakage of $8.154 \times 10^{-21}$. Supplementary Note S4 details these checks. These tests validate the gate-level ROT and TRANS transformations at the full spatial resolution and selector width used in the benchmark.

**Full-scale pose-overlap validation.** The full-scale FFT-derived normalized interaction amplitudes were compared with an independent non-FFT implementation using identical 19-qubit encoded input vectors. For each receptor, occupancy, rotation, and translation index, the reference independently performed the specified discrete rotation and cyclic translation on the ligand occupancy grid coordinates and calculated the resulting overlap by sparse direct summation, including the unchanged auxiliary contribution. Across eight receptor systems, two receptor states, 1,024 ligand occupancies, 32 rotations, and 512 translations per system, all 268,435,456 indexed configurations agreed within absolute and relative tolerances of $10^{-9}$. The maximum absolute difference in $A_0$ was $2.43 \times 10^{-17}$ (overall root mean square error of $2.22 \times 10^{-18}$). Supplementary Note S4 details the comparison. These results establish full-scale numerical agreement for the ROT and TRANS transformations and normalized overlap calculations with the independent direct reference.

**FLAG acceptance across screening thresholds.** Figure 2 shows $p$, the initial FLAG acceptance probability, versus $A_{0,\mathrm{thr}}$ for $N_{\mathrm{OCC}}$ = 1, 16, 128, and 1024. Each panel compares $k$ = 1, 2, 4, and 8. In the deliberately constructed nested benchmark libraries, $p$ generally decreased with more stringent thresholds and larger $N_{\mathrm{OCC}}$.

At $A_{0,\mathrm{thr}} \approx 3.07 \times 10^{-3}$ and $k = 1$, median $p$ was 0.679, 0.122, 0.0702, and 0.0299 for $N_{\mathrm{OCC}}$ = 1, 16, 128, and 1024, respectively. At $N_{\mathrm{OCC}}$ = 1024, increasing $k$ from 1 to 8 raised $p$ from 0.0299 to 0.0739, corresponding to expected FLAG acceptance percentages before QSEL of approximately 3.0% and 7.4%. At $A_{0,\mathrm{thr}} \approx 7.49 \times 10^{-7}$ and $k = 1$, $p$ decreased from

$2.96 \times 10^{-4}$ at $N_{\mathrm{OCC}} = 1$ to $3.69 \times 10^{-7}$ at $N_{\mathrm{OCC}} = 1024$.

At the intermediate threshold $A_{0,\mathrm{thr}} \approx 1.92 \times 10^{-4}$, median $p$ at $k = 1$ was 0.181 for $N_{\mathrm{OCC}} = 1$ and $7.06 \times 10^{-4}$ for $N_{\mathrm{OCC}} = 1024$. With $k = 8$, the corresponding values were 0.244 and $1.06 \times 10^{-3}$. Increasing evaluation resolution raised $p$ for both the smallest and largest tested libraries, although acceptance probabilities were substantially lower for the larger library.

Supplementary Figure S2 shows the qualifying fraction $q$ calculated from the underlying grid scores. Unlike $p$, $q$ does not depend on $k$.

At the most stringent threshold, the qualifying count remained nearly unchanged across the constructed libraries despite a 1,024-fold difference in their total configuration counts (Supplementary Figure S2).

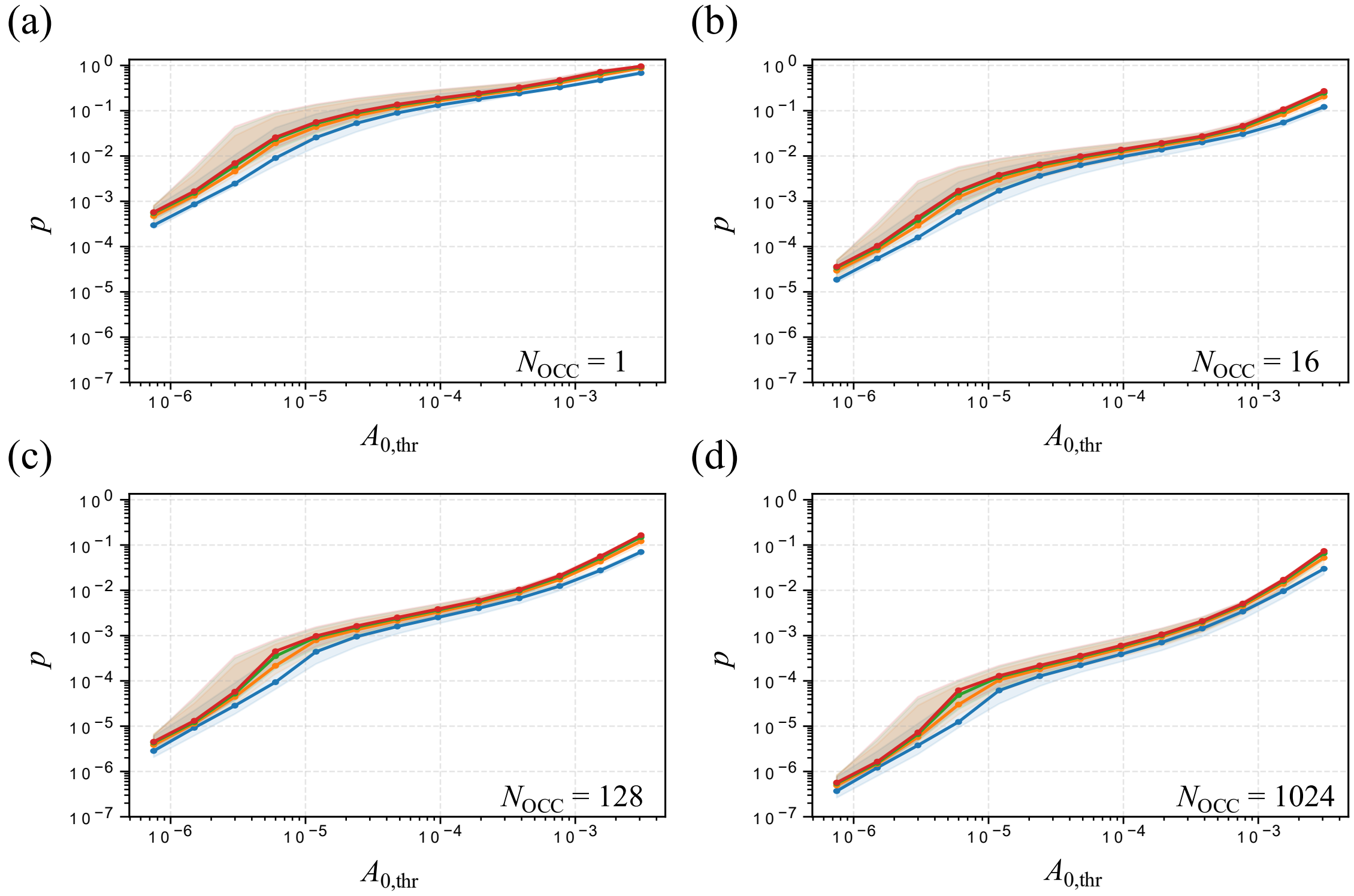

**Figure 2.** Expected FLAG-selected fraction across screening thresholds. Panels (a)–(d) show $p$ versus $A_{0,\mathrm{thr}}$ for $N_{\mathrm{OCC}} = 1$, 16, 128, and 1024, respectively. Blue, orange, green, and red solid curves indicate $k = 1$, 2, 4, and 8, respectively. Curves show medians across eight receptors. Shaded bands indicate the 25th–75th percentiles (IQRs).

**Reliability of FLAG-selected candidate indices.** Figure 3 shows the classification precision $\eta$ of FLAG-selected indices versus $A_{0,\mathrm{thr}}$ for four library sizes and $k = 1, 2, 4$, and 8. Here $\eta$ is evaluated against the benchmark scores and represents the expected fraction of selected indices that satisfy the screening threshold under each tested condition. Increasing $k$ generally improved $\eta$, although the size of the improvement depended on the library and threshold. At $N_{\mathrm{OCC}} = 1024$ and $A_{0,\mathrm{thr}} \approx 3.07 \times 10{-}3$, median $\eta$ rose from 0.628 at $k = 1$ to 0.935 at $k = 8$. At $A_{0,\mathrm{thr}} \approx 7.49 \times 10{-}7$, it increased from 0.512 to 0.907.

The library-size dependence was also evident at the least stringent threshold. At $A_{0,\mathrm{thr}} \approx 3.07 \times 10^{-3}$, median $\eta$ at $k = 1$ decreased from approximately 1.000 for $N_{\mathrm{OCC}} = 1$ to 0.628 for $N_{\mathrm{OCC}} = 1024$. The corresponding medians at $k = 8$ were 0.999 and 0.935. At the intermediate threshold $A_{0,\mathrm{thr}} \approx 1.92 \times 10^{-4}$ and $N_{\mathrm{OCC}} = 1024$, increasing $k$ from 1 to 8 raised median $\eta$ from 0.819 to 0.962.

At the most stringent tested threshold, receptor-specific classification precision varied across the eight systems. For $N_{\mathrm{OCC}} = 1024$ and $A_{0,\mathrm{thr}} \approx 7.49 \times 10^{-7}$, the 25th–75th percentile interval for $\eta$ was 0.343–0.636 at $k = 1$, compared with 0.836–0.930 at $k = 8$. The higher-resolution interval was narrower and shifted upward, indicating improved classification precision and less dispersion among the central half of receptor-specific results.

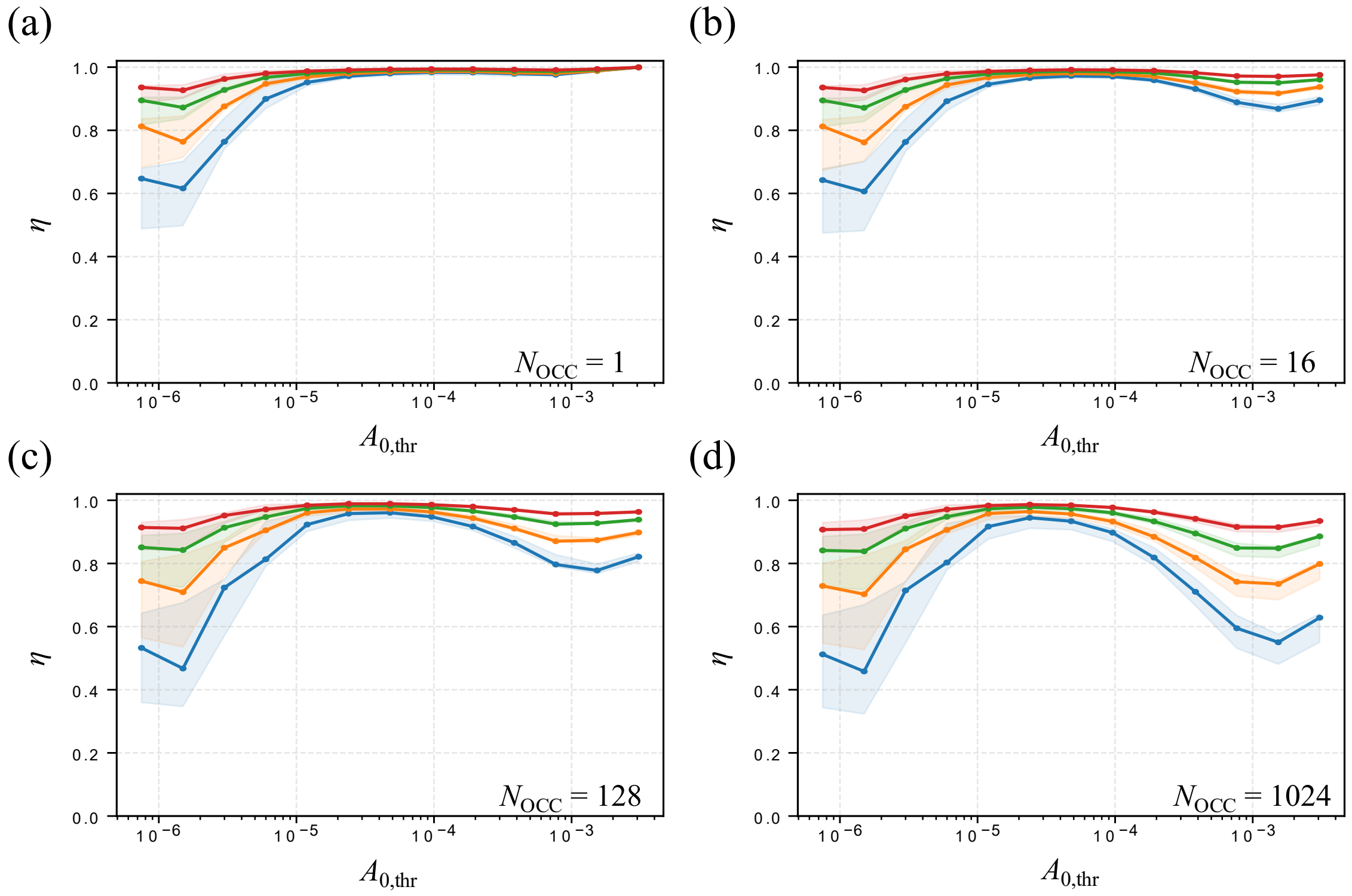

**Figure 3.** Classification precision of FLAG-selected indices across screening thresholds. Panels (a)–(d) show $\eta$ at $N_{\mathrm{OCC}} = 1$, 16, 128, and 1024, respectively, versus $A_{0,\mathrm{thr}}$. Blue, orange, green, and red solid curves correspond to $k = 1$, 2, 4, and 8, respectively. Curves show medians across eight receptors. Shaded bands indicate the 25th–75th percentiles (IQRs).

**QSEL iterations and required shots.** Figure 4 reports $S_{95}(r)$ versus $r$ at $N_{\mathrm{OCC}} = 1024$ for $m_0 = 10$, 14, 18, and 22. Each panel compares $k = 1$, 2, 4, and 8 and includes $r = 0$ as the baseline without QSEL. At $m_0 = 10$ and $k = 1$, the eight-receptor median decreased from 99 shots at $r = 0$ to 1 shot at $r = 4$. At $m_0 = 14$ and $k = 1$, it decreased from 4,250.5 shots at $r = 0$ to 1 shot at $r = 26$. The half-integer median arises from eight receptor-specific integer shot counts.

The regions requiring fewer shots shifted toward larger $r$ as $m_0$ increased. At $m_0 = 18$ and $k = 1$, the median decreased from 55,851.5 shots at $r = 0$ to 2.5 shots at $r = 73$. At $m_0 = 22$ and $k = 1$, it decreased from 8,134,115.5 shots at $r = 0$ to 2.5 at $r = 911$. In the displayed ranges, several curves flattened or rose again after their low-shot regions.

Increasing the resolution multiplier also produced substantial shot reductions. At $m_0 = 10$ and $k = 8$, median $S_{95}(r)$ decreased from 39.5 shots at $r = 0$ to 1 shot at $r = 2$. At $m_0 = 14$ and $k = 8$, it decreased from 2,813.5 at $r = 0$ to 1 at $r = 21$. At $m_0 = 22$ and $k = 8$, the reported median declined from $5.32 \times 10^6$ at $r = 0$ to 4.5 at $r = 568$. The region requiring fewer shots therefore depended on both the screening threshold and the resolution multiplier.

Supplementary Figure S3 repeats the $S_{95}(r)$ comparison at $m_0 = 14$ for $N_{OCC} = 1$, 16, 128, and 1024. These comparisons show how library size and composition affect the shot requirements within the constructed benchmark.

In Supplementary Figure S3, with $m_0 = 14$ and $k = 1$, the median no-QSEL shot counts were 15.5, 221, 757, and 4,250.5 for $N_{OCC} = 1$, 16, 128, and 1024, respectively. The respective lowest plotted median counts occurred at $r = 1$, 5, 10, and 26. The library-size comparison therefore showed that the region requiring fewer shots shifted toward more QSEL iterations at the same screening threshold and evaluation resolution.

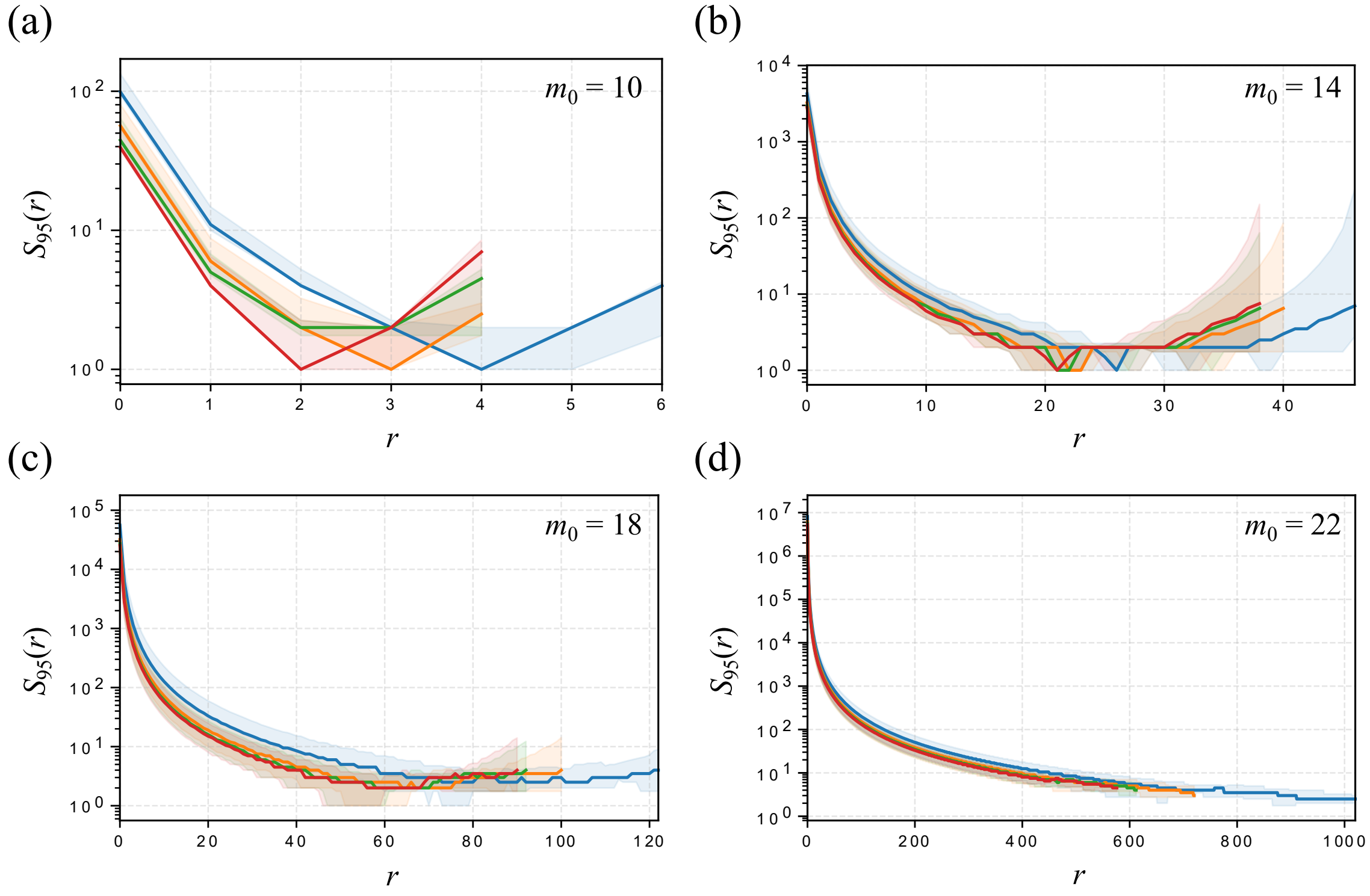

**Figure 4.** Shots $S_{95}(r)$ to obtain FLAG = 1 with 95% probability versus $r$. Panels (a)–(d) use $m_0$ = 10, 14, 18, and 22 with $N_{OCC} = 1024$. Blue, orange, green, and red solid curves indicate $k$ = 1, 2, 4, and 8, respectively. Curves show medians across eight receptors. Shaded bands indicate the 25th–75th percentiles (IQRs).

**Logical $Q$-operator calls and QSEL iteration selection.** Figure 5 reports $C_{95}(r)$ for the same four $m_0$ settings, the same four $k$ values, and the same $N_{OCC}$ as Figure 4. The curves combine the calculated shot requirements and the logical $Q$-operator calls per shot. The latter increase with $r$ and $m$.

At $m_0$ = 10 and $k$ = 1, the eight-receptor median $C_{95}(r)$ fell from 101,277 calls at $r$ = 0 to 9,207 calls at $r$ = 4. At $m_0$ = 14 and $k$ = 1, it fell from 6.96 × $10^7$ calls at $r$ = 0 to 8.68 × $10^5$ calls at $r$ = 26. At $m_0$ = 10, the median cost at $r$ = 0 was 323,545 calls for $k$ = 8, compared with 101,277 calls for $k$ = 1. In Figures 5(a)–(c), several curves increased after their low-cost regions. Figure 5(d) covers a larger sampled range of $r$.

The cost reductions also extended to the more stringent settings. At $m_0$ = 18 and $k$ = 1, the median $C_{95}(r)$ decreased from 1.46 × $10^{10}$ calls at $r$ = 0 to 9.63 × $10^7$ at $r$ = 73. At $m_0$ = 22 and $k$ = 1, it decreased from 3.41 × $10^{13}$ calls at $r$ = 0 to 1.91 × $10^{10}$ at $r$ = 911. Although QSEL reduced the plotted logical $Q$-operator call counts, their absolute magnitudes differed substantially across the four threshold settings.

Supplementary Figure S4 shows $C_{95}(r)$ for $N_{OCC}$ = 1, 16, 128, and 1024 at $m_0$ = 14, matching the parameter settings in Supplementary Figure S3. Supplementary Figure S5 summarizes $\min_r C_{95}(r)$ versus $N_{OCC}$ over the eleven nested libraries at $m_0$ = 10, 14, 18, and 22. For each receptor, the minimum is taken over the sampled $r$ values before calculating the median. These supplementary results provide library-size comparisons not included in the main Figures 4 and 5 at fixed NOCC.

Resolution also changed the minimum plotted logical workload. At $m_0$ = 10, the lowest plotted median $C_{95}(r)$ was 9,207 calls for $k$ = 1 at $r$ = 4, compared with 40,955 calls for $k$ = 8 at $r$ = 2. At $m_0$ = 14, the corresponding minima were 868,299 and 5.64 × $10^6$ calls. Supplementary Figure S5 further shows that, after minimizing cost separately for each receptor, the median minimum at $m_0$ = 14 and $k$ = 1 increased from 98,298 calls for $N_{OCC}$ = 1 to 851,916 for $N_{OCC}$ = 1024.

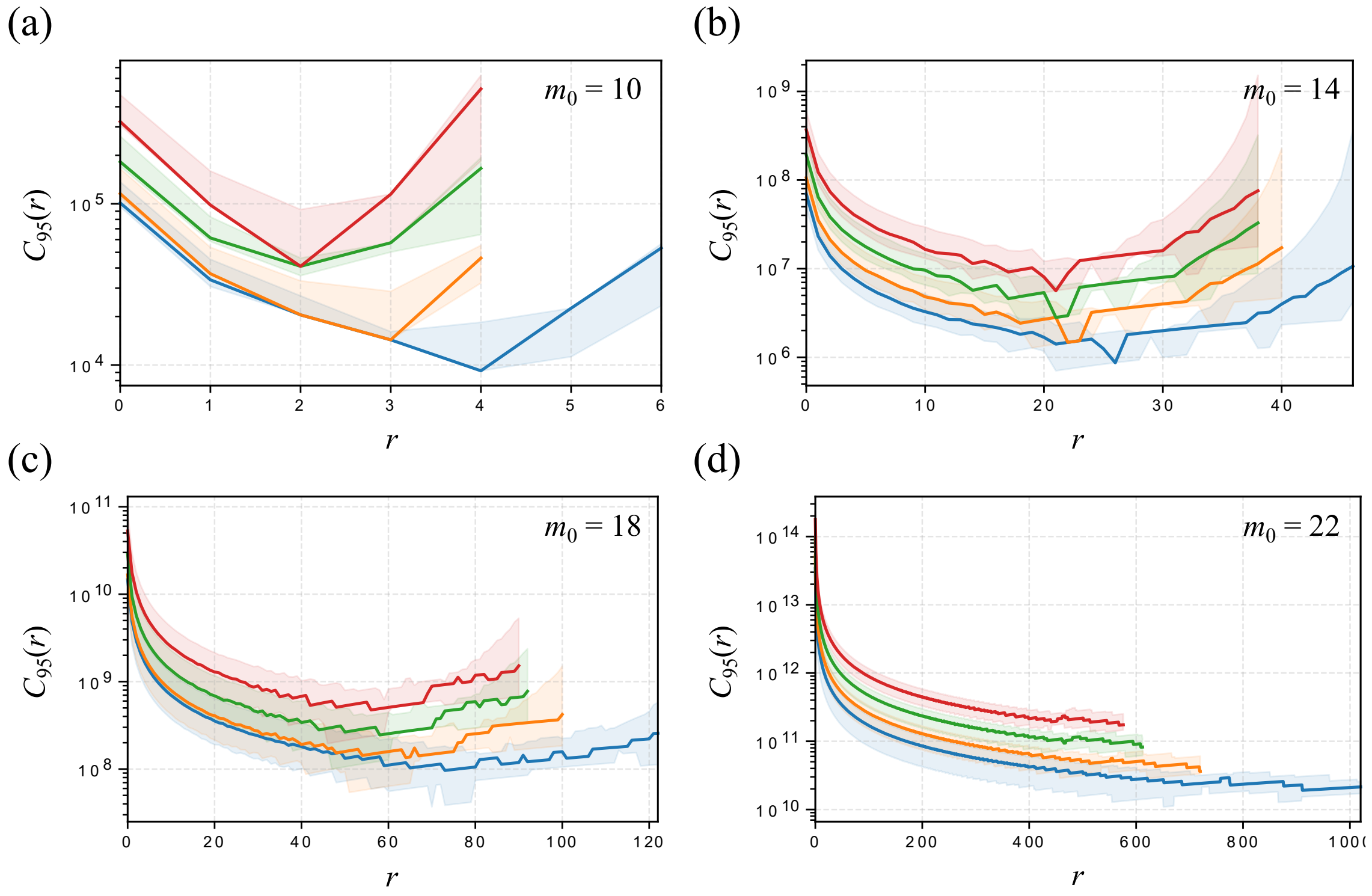

Figure 5. Logical $Q$-operator calls required for FLAG = 1 with 95% probability. Panels (a)–(d) show $C_{95}(r)$ versus QSEL iterations $r$ at $m_0$ = 10, 14, 18, and 22, respectively, with $N_{OCC}$ = 1024. Blue, orange, green, and red solid curves correspond to $k$ = 1, 2, 4, and 8, respectively. Curves show medians across eight receptors. Shaded bands indicate the 25th–75th percentiles (IQRs).

## Discussion

**FLAG-selected ligand and pose indices.** The objective of the present framework is to turn a coherently encoded molecular interaction criterion into a recoverable SBVS decision. Earlier grid-based studies supported numerical interaction-score estimation across indexed configurations. Here, the intended output is instead a candidate configuration index. QAE/QPE and QCOMP write a threshold decision to FLAG. Once FLAG = 1 is measured, the selector registers identify the receptor state, ligand occupancy, rotation, and translation without requiring the numerical scores of all other configurations.

**Screening threshold and evaluation resolution.** The screening threshold $A_{0,\mathrm{thr}}$ defines which normalized grid-interaction amplitudes meet the selection criterion. A lower threshold imposes a more stringent selection criterion and generally reduces the initial FLAG acceptance probability $p$ and increases the number of shots required for FLAG detection without QSEL. In the numerical parameter sweep, lower thresholds correspond to larger $m_0$ and, at fixed $k$, larger $m$, increasing the logical $Q$-operator calls per evaluation.

At a fixed threshold, the QAE/QPE evaluation resolution influences the accuracy of the threshold decision written to FLAG. In the numerical benchmark, increasing the resolution multiplier $k$ adds evaluation qubits according to $m = m_0 + \log_2(k)$. Greater resolution generally improves classification precision $\eta$, making FLAG-selected indices more reliable. For example, at $N_{\mathrm{OCC}} = 1024$ and $A_{0,\mathrm{thr}} \approx 3.07 \times 10^{-3}$, the median $\eta$ rose from 0.628 at $k = 1$ to 0.935 at $k = 8$ (Figure 3). The trade-off is greater logical work because an $m$-qubit QAE/QPE evaluation requires $2^m - 1$ calls to the internal $Q$-operator.

**Library size and pose representation.** The occupancy count $N_{\mathrm{OCC}}$ controls the number of indexed ligand occupancies. Doubling $N_{\mathrm{OCC}}$ adds one occupancy-selector qubit in the nested libraries, while preparing and selecting a larger collection of ligand occupancy states can also increase the complexity of the coherent input circuit. The present pose representation uses five ROT-selector qubits and nine TRANS-selector qubits, and the normalized receptor–ligand overlap uses a fixed 19-qubit register comprising DATA and AUX. The seven selected AutoDock atom types and the 32 × 32 × 32 spatial grid are benchmark settings. Supplementary Note S1 generalizes the inner-product representation to additional atom-type affinity maps, a separate charge-dependent desolvation map, and ligand-dependent non-grid scalar terms. These additional contributions were not evaluated in the eight-channel benchmark and would require corresponding data encoding and circuit resource accounting. Larger map boxes would also require extended coordinate registers and controlled pose operations.

Library size alone, however, does not determine the shots required to retrieve a FLAG-selected index. The relevant quantity is the initial FLAG acceptance probability $p$. Acceptance may remain similar if the number of accepted configurations increases proportionally with library size or may decrease if it grows more slowly. Figure 2 shows that $p$ decreases with increasing $N_{\mathrm{OCC}}$ in the tested libraries.

**QSEL iterations and FLAG-detection shots.** QSEL increases the probability of measuring FLAG = 1 by coherently amplifying the FLAG-selected subspace. If $p$ is small, an appropriate number $r$ of iterations can substantially reduce $S_{95}(r)$. The first useful amplification peak depends on $p$ and therefore on the threshold and library. Because amplification oscillates, excessive $r$ can raise the shot requirement again. Ideal QSEL preserves classification precision $\eta$. It increases the probability of obtaining an accepted index without improving the reliability of the accepted indices.

The effect is visible in Figure 4. At $N_{\mathrm{OCC}} = 1024$, $m_0 = 10$, and $k = 1$, the median $S_{95}(r)$ decreased from 99 shots without QSEL to one shot at $r = 4$. At $m_0 = 14$ and $k = 1$, the corresponding median decreased from 4,250.5 shots at $r = 0$ to one shot at $r = 26$. The larger iteration count reflects the lower initial acceptance probability. The choice of $r$ therefore depends on $p$ and the chosen cost objective.

**Balancing shots and logical work.** The number of shots is only one component of the retrieval workload. Each QSEL iteration introduces an additional forward and inverse QAE/QPE evaluation block, so a shot after $r$ iterations contains $2r + 1$ evaluation blocks. Since each block uses $2^m - 1$ calls to the internal $Q$-operator, $C_{95}(r)$ incorporates both the number of shots and the additional operations performed within each shot. Minimizing $S_{95}(r)$ does not necessarily minimize $C_{95}(r)$. Figures 4 and 5 should therefore be considered together.

For example, at $m_0 = 10$ and $k = 1$, QSEL reduced the median shot count from 99 to one at $r = 4$, but the median logical cost decreased from 101,277 to 9,207 $Q$-operator calls. This represents an approximately elevenfold reduction rather than a 99-fold reduction. Moreover, the minimum plotted cost at $m_0 = 10$ was 9,207 calls for $k = 1$, compared with 40,955 calls for $k = 8$. Higher resolution can improve $\eta$, but costlier evaluations can offset shot reductions. The choice must balance the screening threshold, classification precision, QSEL iterations, and logical workload.

**Logical *Q*-operator calls and physical gate costs.** The cost $C_{95}(r)$ counts applications of the internal Grover operator used in QAE/QPE. It does not count elementary quantum gates, circuit depth, or runtime. Each call includes overlap preparation, its inverse, and reflections, so its gate-level complexity depends on data loading, selector-controlled ROT/TRANS, and other subcircuits. More occupancy, pose-selector, or DATA/AUX qubits can raise the gate cost per call without proportionally changing $C_{95}(r)$. Increasing $m$ raises the number of calls per evaluation block. Increasing $r$ raises the number of blocks per shot, although the total $C_{95}(r)$ also depends on the resulting shot count. The 23-qubit and 29-qubit Qiskit tests validate the implemented circuit functions, whereas Figures 4 and 5 report logical $Q$-operator call counts. Physical gate counts and runtime require implementation-specific accounting [43].

**Directions for further development.** The central direction for further development is to maintain a modular separation between molecular evaluation and threshold-qualified index retrieval. In the present implementation, AutoDock grid interactions provide the evaluation model, while QAE/QPE, QCOMP, and QSEL perform coherent threshold classification and candidate index retrieval. This organization allows alternative molecular evaluation methods to be incorporated without changing the basic index-first retrieval objective, provided that their outputs can be coherently encoded for threshold comparison. A major extension is to develop more physically complete evaluation models that incorporate quantum electronic-structure calculations [44-46], improved solvation treatments [47, 48], and conformational entropy [1-3] toward binding free-energy assessment. Quantum algorithms for solving the Poisson equation [49, 50] may provide a mathematical basis for future electrostatic modeling, although their extension to realistic solvation calculations requires further development. Because binding free energy depends on ensembles of molecular states, this development requires suitable coherent formulations of the relevant thermodynamic quantities. Another direction is validation on physical quantum computers.

Reduced-scale circuits can be designed according to the available qubit counts, connectivity, native gate sets, coherence times, and noise characteristics of specific devices. Hardware-adapted implementations would enable experimental tests of threshold classification and candidate index retrieval, while guiding subsequent improvements to the architecture.

## Conclusion

This study extends our earlier quantum SBVS studies from molecular representation and numerical interaction-score estimation to threshold-qualified ligand and pose index retrieval. In the proposed coherent architecture, QAE/QPE and QCOMP produce a FLAG decision, and optional QSEL amplifies the accepted subspace. The index registers identify a candidate molecular configuration without requiring complete numerical score readout. In the eight-receptor AutoDock grid benchmark, the initial FLAG acceptance probability $p$ and classification precision $\eta$ characterized candidate yield and reliability. The shot budget $S95(r)$ and logical $Q$-operator call count $C95(r)$ quantified the measurement and coherent-work trade-offs. Higher QAE/QPE resolution generally improved classification precision at additional evaluation cost, whereas QSEL could reduce required shots without necessarily minimizing logical work. A reduced 23-qubit integrated circuit and a 29-qubit full-resolution ROT/TRANS subcircuit were validated separately, while the large-library results used an ideal logical probability model. These results establish the behavior and limitations of an index-oriented retrieval strategy for SBVS. The AutoDock grid model serves as a demonstrator of the architecture. Replacing it with more physically detailed, coherently implementable energy and free-energy estimators remains a longer-term direction for exploring large molecular configuration spaces.

## Supporting Information

Supplementary Note S1 details the generalized AutoDock4 grid-channel and eligible non-grid scoring representation and its normalized inner-product encoding. Supplementary Notes S2–S7 describe the general FLAG classification and cost model, parameter settings, 23-qubit and 29-qubit gate-level validations, full-scale numerical overlap checks, and additional analyses. Figure S1 depicts the 23-qubit validation circuit. Figure S2 reports $q$. Figures S3 and S4 compare $S_{95}(r)$ and $C_{95}(r)$ across library sizes. Figure S5 reports $\min_r C_{95}(r)$. Tables S1 and S2 provide selector register and evaluation resolution settings, and Table S3 summarizes the gate-level comparisons.

## Funding

This research did not receive any specific grant from funding agencies in the public, commercial, or not-for-profit sectors.

## Author contributions

P.-K. Yang conceived the study, developed the methodology, performed the simulations and analyses, and wrote the manuscript.

## Competing interests

The author declares no competing financial interest.

## Acknowledgments

The author used ChatGPT (OpenAI) for English-language editing and to improve the clarity and readability of the manuscript.

## Data and Software Availability

The source protein–ligand structures were obtained from the PDBbind database (https://www.pdbbind-plus.org.cn/). Receptor selection and construction of the nested ligand libraries are described in the Methods. The code supporting this study is publicly available at https://github.com/peikunyang/27_QSBVS

**Supporting Information for Quantum Amplitude Estimation for Ligand and Pose Index Retrieval**

Pei-Kun Yang

E-mail: peikun@isu.edu.tw, peikun6416@gmail.com

**Supplementary Note S1. Inner-product formulation of grid-based protein–ligand scoring**

The protein–ligand scoring expression is represented as a sum of products between receptor-dependent grid maps and ligand-dependent distributions on a common Cartesian grid. The construction accommodates the electrostatic map, the atom-type affinity maps required by the chosen AutoDock4 parameterization, and the desolvation map. A configuration-dependent term that is not represented by a receptor grid can also be incorporated as a separate scalar channel when it depends only on the ligand configuration. All contributions retain the scoring parameters and weights of the selected model. Equations (S1.1)–(S1.6) give illustrative expressions for a fixed receptor and ligand configuration, with configuration indices omitted. The indexed formulation begins with Equation (S1.7).

**Electrostatic interaction.** The unscreened atom-pair Coulomb expression provides a reference for the electrostatic term. For receptor atoms $i$ and ligand atoms $j$, it is

$$4\pi\varepsilon_0 E_{\mathrm{ele}} = \sum_i \sum_j \frac{Q_i q_j}{\left|\boldsymbol{r}_i - \boldsymbol{r}_j\right|} \tag{S1.1}$$

Here $Q_i$ and $q_j$ are the partial charges, $\boldsymbol{r}_i$ and $\boldsymbol{r}_j$ are the atomic position vectors, and $\varepsilon_0$ is the vacuum permittivity. For a positive elementary test charge $e$ placed at grid point $g$, the corresponding reference potential-energy map is

$$U_{\mathrm{e},g} = \frac{e}{4\pi\varepsilon_0} \sum_i \frac{Q_i}{\left|\boldsymbol{r}_i - \boldsymbol{r}_g\right|} \tag{S1.2}$$

The AutoGrid4 electrostatic map can use a distance-dependent dielectric and an empirical electrostatic weight, rather than the unscreened Coulomb expressions in Equations (S1.1) and (S1.2). For the fixed configuration considered here, let $U_{e,g}$ denote the resulting electrostatic map value at grid point $g$. Trilinear accumulation of ligand partial charges, with their signs retained, defines $N_{e,g}$, the interpolated ligand charge distribution expressed in units of the elementary charge. The grid approximation is

$$E_{\mathrm{ele}} \approx \sum_{g=1}^{n_g} N_{\mathrm{e},g} U_{\mathrm{e},g} \tag{S1.3}$$

Here $n_g$ denotes the number of grid points. The approximation reflects spatial discretization and interpolation, whereas the electrostatic interaction model is determined by the selected AutoGrid4 settings.

**Atom-type affinity maps.** The van der Waals (vdW) contribution can be illustrated by the Lennard–Jones 12–6 expression

$$E_{\mathrm{vdW}} = \sum_i \sum_j \sqrt{\varepsilon_i \varepsilon_j} \left[ \left( \frac{R_{\min,i} + R_{\min,j}}{\left|\boldsymbol{r}_i - \boldsymbol{r}_j\right|} \right)^{12} - 2\left( \frac{R_{\min,i} + R_{\min,j}}{\left|\boldsymbol{r}_i - \boldsymbol{r}_j\right|} \right)^{6} \right] \tag{S1.4}$$

The positive parameters $\varepsilon_i$ and $\varepsilon_j$ are atomic well-depth magnitudes, $R_{\min,i}$ and $R_{\min,j}$ are the half-equilibrium-distance parameters, and $r_{ij} = \|\boldsymbol{r}_i - \boldsymbol{r}_j\|$. The geometric mean specifies the pair well depth. The twelfth- and sixth-power terms describe repulsion and attraction, respectively.

For a probe atom of type $t$ at grid point $g$, the corresponding illustrative vdW map is

$$U_{\mathrm{v},t,g} = \sum_i \sqrt{\varepsilon_i \varepsilon_t} \left[ \left( \frac{R_{\min,i} + R_{\min,t}}{\left| \boldsymbol{r}_i - \boldsymbol{r}_g \right|} \right)^{12} - 2 \left( \frac{R_{\min,i} + R_{\min,t}}{\left| \boldsymbol{r}_i - \boldsymbol{r}_g \right|} \right)^{6} \right] \tag{S1.5}$$

Here $r_{ig} = \|r_i - r_g\|$. The dimensionless occupancy $N_{v,t,g}$ represents the trilinearly assigned occupancy of ligand atoms of type $t$ at grid point $g$. The corresponding illustrative vdW energy is

$$E_{\mathrm{vdW},t} \approx \sum_{g=1}^{n_g} N_{\mathrm{v},t,g} U_{\mathrm{v},t,g} \tag{S1.6}$$

The complete AutoGrid4 atom-type affinity map $U_{t,g}$ need not equal the isolated vdW component in Equation (S1.5). Depending on the ligand atom type and the scoring parameters, an affinity map can also contain hydrogen-bond interactions, including the AutoDock4 12–10 contribution. In the usual combined-map representation, it also contains the charge-independent desolvation contribution. The set $T$ denotes the ligand atom types for which affinity maps are generated. Its size is determined by the chosen parameterization rather than a fixed number of types.

**Desolvation and general scoring expression.** Let $rc$ label a receptor state, and let $c = (lt, lc, ro, tr)$ denote ligand identity, conformation, rotation, and translation. AutoGrid4 also supplies the electrostatic map (e.map) and a charge-dependent desolvation map (d.map). Let $U_{\mathrm{d},g}^{\mathrm{rc}}$ denote the latter map and $N_{\mathrm{d},g}^{c}$ its corresponding interpolated ligand distribution, weighted by the magnitudes of ligand partial charges according to the AutoDock4 desolvation convention. For the standard combined-map representation, define the spatial channel set $A = \{\mathrm{e}, \mathrm{d}\} \cup T$. The general grid-based scoring expression is

$$E^{\mathrm{rc,c}} = \sum_{\alpha \in A} \sum_{g=1}^{n_g} N_{\alpha,g}^{\mathrm{c}} U_{\alpha,g}^{\mathrm{rc}} + E_{\mathrm{nonmap}}^{\mathrm{c}} \tag{S1.7}$$

In Equation (S1.7), the channel index $\alpha$ covers electrostatic (e), atom-type affinity ($t \in T$), and charge-dependent desolvation (d) contributions. When AutoGrid4 generates that contribution as separate maps, the channel set $A$ can instead be extended to include those maps and their matching ligand weights. The non-grid scalar $E_{\mathrm{nonmap}}^{\mathrm{c}}$ includes terms such as the torsional penalty and any ligand-only intramolecular or unbound-state corrections specified by the scoring protocol. The scalar-channel representation below assumes that this term depends only on $c$ and not on $rc$. Non-grid corrections outside this decomposition require separate configuration-controlled evaluation.

**Normalized inner-product representation.** Using the indexed quantities and channel set $A$ defined above, choose a common energy offset $E^+$ so that $E^{rc,c} + E^+ \geq 0$ over the indexed configurations. The scoring energy is expressed as

$$E^{rc,c} = L_{\mathrm{U}} L_{\mathrm{N}} \boldsymbol{U}^{rc} \bullet \boldsymbol{N}^{c} - E^{+} \tag{S1.8}$$

Let $\boldsymbol{u}^{rc}$ and $\boldsymbol{n}^{c}$ be the concatenated map and ligand-weight vectors over all channel labels $\alpha \in A$ and grid points $g$. Their entries are $U_{\alpha,g}^{rc}$ and $N_{\alpha,g}^{c}$, respectively, using the same channel ordering. The normalized receptor-state vector is

$$\boldsymbol{U}^{rc} = \frac{1}{L_{\mathrm{U}}}\begin{pmatrix}\boldsymbol{u}^{rc} & E_{\mathrm{ref}} & E^{+} & x_{\mathrm{U}}^{rc} & 0\end{pmatrix} \tag{S1.9}$$

The corresponding ligand-configuration vector is

$$\boldsymbol{N}^{c} = \frac{1}{L_{\mathrm{N}}}\begin{pmatrix}\boldsymbol{n}^{c} & E_{\mathrm{nonmap}}^{c}/E_{\mathrm{ref}} & 1 & 0 & x_{\mathrm{N}}^{c}\end{pmatrix} \tag{S1.10}$$

Here $E_{\mathrm{ref}} > 0$ is a fixed reference energy for the non-grid scalar channel. Each vector combines grid channels with scalar, offset, and orthogonal padding components. The selector superscripts are indices, not powers. The Hilbert-space register accommodates the channels and auxiliary basis states.

**Common normalization and padding.** To maintain the same normalization across receptor states within a system, select $L_U > 0$ satisfying

$$\left(L_{\mathrm{U}}\right)^2 \geq \max_{\mathrm{rc}}\left(\left\|\boldsymbol{u}^{rc}\right\|^2 + \left(E_{\mathrm{ref}}\right)^2 + \left(E^{+}\right)^2\right) \tag{S1.11}$$

The receptor-side padding amplitude is

$$x_{\mathrm{U}}^{rc} = \sqrt{\left(L_{\mathrm{U}}\right)^2 - \left\|\boldsymbol{U}^{rc}\right\|^2 - \left(E_{\mathrm{ref}}\right)^2 - \left(E^{+}\right)^2} \tag{S1.12}$$

A common ligand normalization factor $L_N > 0$ is chosen to satisfy

$$\left(L_{\mathrm{N}}\right)^2 \geq \max_{\mathrm{c}}\left(\left\|\boldsymbol{n}^{c}\right\|^2 + \left(E_{\mathrm{nonmap}}^{c}/E_{\mathrm{ref}}\right)^2 + 1\right) \tag{S1.13}$$

The ligand-side padding amplitude is

$$x_{\mathrm{N}}^{c} = \sqrt{\left(L_{\mathrm{N}}\right)^2 - \left\|\boldsymbol{n}^{c}\right\|^2 - \left(E_{\mathrm{nonmap}}^{c}/E_{\mathrm{ref}}\right)^2 - 1} \tag{S1.14}$$

The radicands in Equations (S1.12) and (S1.14) are nonnegative by construction. Consequently, $\|\boldsymbol{U}^{rc}\| = \|\boldsymbol{N}^{c}\| = 1$, and $0 \leq \boldsymbol{U}^{rc} \cdot \boldsymbol{N}^{c} \leq 1$ under the selected nonnegative energy shift. When a chosen scoring model omits the non-grid contribution, its scalar channel can be omitted. Likewise, unused map channels are absent from the corresponding vectors. Thus restricted scoring models are special cases of this general construction, while channel count, auxiliary-space dimension, and circuit resources depend on the implemented terms.

**Supplementary Note S2. Threshold-classification probabilities and logical retrieval costs**

**Interaction amplitude and threshold definition.** Using the vectors in Supplementary Note S1, the normalized interaction amplitude is $A_0(rc, c) = \boldsymbol{U}^{rc} \cdot \boldsymbol{N}^{c} = (E^{rc,c} + E^{+})/(L_{\mathrm{U}}\, L_{\mathrm{N}})$, where $rc$ denotes a receptor state and $c = (lt, lc, ro, tr)$. These indices specify a receptor state and ligand configuration, with $lc$ fixed in the benchmark. The offset and normalization ensure $0 \leq A_0(rc, c) \leq 1$.

The threshold-qualified event $G$ is $A_0(rc, c) < A_{0,\mathrm{thr}}$, where $A_{0,\mathrm{thr}}$ is the screening threshold. Quantum amplitude estimation (QAE) based on quantum phase estimation (QPE) supplies finite-precision phase information to a reversible quantum comparator (QCOMP). The low- and high-phase regions follow the same QPE phase convention described in the main Methods. QCOMP writes its threshold decision to the binary FLAG register. When FLAG = 1 is observed, measuring the selector registers returns a candidate ligand–pose index. With finite phase resolution, a FLAG-selected index need not meet the threshold. The benchmark classification precision $\eta$ quantifies the proportion of selected indices that satisfy the threshold.

Let $N_{\mathrm{OCC}}$ denote the number of ligand occupancies. The indexed search contains two receptor states, 32 rotation selector states, and 512 cyclic translations, giving $N_{\mathrm{search}}$ = 32,768 $N_{\mathrm{OCC}}$ configurations. The QAE/QPE evaluation register contains $m$ qubits, and QCOMP uses the screening threshold $A_{0,\mathrm{thr}}$. For the numerical study, we set the screening threshold to $A_{0,\mathrm{thr}} = \sin(\pi/2^{m_0})$, corresponding to the adopted near-zero amplitude-resolution step at the base setting $m_0$. We then set $m = m_0 + \log_2(k)$, with $k$ = 1, 2, 4, and 8, to compare different evaluation resolutions at the same threshold. We used $m_0$ = 10–22. The resulting $A_{0,\mathrm{thr}}$ values were approximately $3.07 \times 10^{-3}$, $1.92 \times 10^{-4}$, $1.20 \times 10^{-5}$, and $7.49 \times 10^{-7}$ for $m_0$ = 10, 14, 18, and 22, respectively.

**Unitary operators for QAE/QPE and quantum selection (QSEL).** Let $j = (rc, c)$ denote the complete selector index, combining the receptor state $rc$ and ligand–pose configuration $c$. Let $|0^{\mathrm{D}}\rangle$ be the all-zero state of the 19-qubit DATA/AUX register. For a fixed selector state, define $P_{\mathrm{N},c}$ as the unitary that prepares the selected normalized ligand vector, including SELECT-OCC and the rotation (ROT) and translation (TRANS) operations, and $P_{\mathrm{U},rc}$ as the unitary that prepares the selected normalized receptor-map vector. Thus, $P_{\mathrm{N},c}|0^{\mathrm{D}}\rangle = |N^{c}\rangle$ and $P_{\mathrm{U},rc}|0^{\mathrm{D}}\rangle = |U^{rc}\rangle$. The pose operations leave the auxiliary components unchanged.

At the ideal logical-operator level, the SELECT-MAP overlap step applies the inverse of receptor-map preparation after the ligand state has been prepared and posed. Therefore $W_j = P_{\mathrm{U},rc}{}^{\dagger} P_{\mathrm{N},c}$, and $\langle 0^{\mathrm{D}}|W_j|0^{\mathrm{D}}\rangle = \langle U^{rc}|N^{c}\rangle = A_{0,j}$. This is the normalized inner product defined in Supplementary Note S1. In particular, $A_{0,j}$ is the amplitude of the DATA/AUX all-zero basis state in $W_j|0^{\mathrm{D}}\rangle$.

For each fixed selector configuration, the internal amplitude-estimation Grover operator $Q_j$ is constructed from $W_j$, $W_j^\dagger$, and the zero-input reflection $S_0$, as defined in Equation (S2.1).

$$Q_j = -W_j S_0 W_j^\dagger S_0 \tag{S2.1}$$

In Equation (S2.1), $S_0$ inverts the phase of the DATA/AUX all-zero basis state. The operator $W_j S_0 W_j^\dagger$ flips the phase of $W_j|0^D\rangle$. With the leading minus sign, the Grover operator $Q_j$ has two eigenphases given by $\varphi_{j,\pm} = \pm\sin^{-1}(A_{0,j})/\pi \pmod 1$.

Across the uniform selector superposition prepared by the initial Hadamard gates, the selector-preserving operator $Q_{AE}$ is constructed from these configuration-dependent Grover operators, as given in Equation (S2.2), where $\Pi_j = |j\rangle\langle j|$ projects onto selector basis state $j$.

$$Q_{AE} = \sum_j \Pi_j \otimes Q_j \tag{S2.2}$$

QPE applies evaluation-register-controlled powers of $Q_{AE}$ and an inverse quantum Fourier transform. The $m$-qubit evaluation register coherently encodes an integer phase estimate $y$ between 0 and $2^m - 1$, representing the eigenphase approximately as $y/2^m$. The ideal QPE probability for integer phase-bin outcome $y$ from an eigenphase $\varphi$ is [1]

$$P_m(y;\phi) = \frac{\sin^2\left[\pi\left(2^m\phi - y\right)\right]}{2^{2m}\sin^2\left[\pi\left(\phi - y/2^m\right)\right]} \tag{S2.3}$$

The expression in Equation (S2.3) is evaluated by its continuous limit when the denominator vanishes. The two eigenphase branches $\varphi_{j,+}$ and $\varphi_{j,-}$ enter with equal weight. At $m = m_0 + \log_2(k)$, QCOMP accepts bins $y < k$ or $y \geq 2^m - k + 1$, corresponding to the threshold $A_{0,\mathrm{thr}} = \sin(\pi/2^{m_0})$ used here. Summing the two-branch distribution over these accepted bins gives the FLAG probability for each configuration. Uniform averaging gives $p$, and $\eta$ is obtained by restricting the FLAG probability weight to configurations satisfying $G$.

QCOMP performs these two disjoint integer comparisons coherently on the evaluation register. Each condition reversibly toggles the initially zero FLAG qubit. Because the accepted ranges do not overlap, the two controlled toggles implement their logical OR. Thus, QCOMP writes FLAG = 1 for an accepted phase bin and FLAG = 0 otherwise, without measuring either the evaluation register or the configuration selector.

After coherent FLAG generation, let $B$ denote the complete unitary circuit that prepares the FLAG-encoded state from the all-zero input state. The QSEL Grover operator is defined in Equation (S2.4).

$$Q_{SEL} = -BS_0B^\dagger S_F \tag{S2.4}$$

In Equation (S2.4), $S_F$ flips the phase of the FLAG = 1 subspace, whereas the zero-input reflection $S_0$ acts on the complete register space used by $B$. This $S_0$ acts on a different register space from the inner DATA/AUX reflection in Equation (S2.1). The action of $B^\dagger$ reverses the coherent preparation, and QSEL amplifies the FLAG = 1 subspace.

**FLAG amplification.** Let $p$ = Pr(FLAG = 1 | $r$ = 0) be the initial FLAG-detection probability after QCOMP. After $r$ nonnegative QSEL Grover iterations, the ideal probability of FLAG = 1 is given by Equation (S2.5). This is Grover amplitude amplification of the FLAG-selected subspace. [2, 3]

$$p_r = P(\mathrm{FLAG}=1\,|\,r) = \sin^2\left[(2r+1)\theta\right] \tag{S2.5}$$

Here $p_r$ is the FLAG-detection probability after $r$ QSEL iterations, and $\theta = \sin^{-1}\sqrt{p}$ is the Grover angle. The first-peak iteration count $r_{\mathrm{peak}}$ lies near the continuous maximum at $(2r + 1)\theta = \pi/2$, giving

$$r_{\mathrm{peak}} \approx \frac{\pi}{4\sin^{-1}\sqrt{p}} - \frac{1}{2} \approx \frac{\pi}{4\sqrt{p}} - \frac{1}{2} \tag{S2.6}$$

The second approximation in Equation (S2.6) applies when $0 < p \ll 1$.

**Shots for FLAG detection.** For $S$ independent executions at fixed $r$, the probability of at least one FLAG = 1 outcome is $1 - (1 - p_r)^S$. Requiring this probability to be at least 0.95 gives

$$S_{95}(r) = \left\lceil \frac{\ln(0.05)}{\ln(1-p_r)} \right\rceil \tag{S2.7}$$

$S_{95}(r)$ is the minimum integer shot count for FLAG detection at 95% probability. The ceiling brackets denote upward rounding, and ln denotes the natural logarithm. $S_{95}(r) = 1$ when $p_r = 1$, while no finite shot count suffices when $p_r = 0$.

**Logical *Q*-operator costs.** The internal amplitude-estimation operator $Q_{\mathrm{AE}}$ defined in Equation (S2.2) is distinct from the outer QSEL operator. In QAE/QPE, the controlled powers of $Q_{\mathrm{AE}}$ require $1 + 2 + 4 + \cdots + 2^{(m-1)} = 2^m - 1$ calls per evaluation block. The cost model includes one initial block and one forward and one inverse block per QSEL iteration, giving $2r + 1$ blocks per shot. Each application of $Q_{\mathrm{AE}}$ or its inverse counts as one call. The total FLAG-detection cost is

$$C_{95}(r) = S_{95}(r)(2r+1)(2^m-1) \tag{S2.8}$$

$C_{95}(r)$ counts logical $Q$-operator calls for FLAG detection at 95% probability. For $p \ll 1$, $S_{95}(0) \approx \ln(20)/p$, giving $C_{95}(0) \approx (2^m - 1)\ln(20)/p$. These counts represent calls to the internal QAE operator rather than elementary-gate counts. The number of gates per call depends on the chosen data-loading, pose-transformation, overlap, and reflection circuits.

For $p \ll 1$, $r_{\mathrm{peak}} = \mathrm{O}(1/\sqrt{p})$ and $S_{95}(r_{\mathrm{peak}}) = \mathrm{O}(1)$. At fixed $m$, $C_{95}(0) = \mathrm{O}(1/p)$ and $C_{95}(r_{\mathrm{peak}}) = \mathrm{O}(1/\sqrt{p})$. Therefore, $C_{95}(r_{\mathrm{peak}})/C_{95}(0) = \mathrm{O}(\sqrt{p})$.

**Classification precision.** Let $M$ be the number of configurations satisfying $G$. Under uniform selector preparation, their qualifying fraction is $q = M/N_{\mathrm{search}} = \Pr(G)$. The correct-marking rate $a$ = Pr(FLAG = 1 | $G$) and false-positive marking rate $b$ = Pr(FLAG = 1 | ¬$G$) are evaluated before QSEL. Thus, $p = qa + (1 - q)b$, and classification precision $\eta$ is given by Bayes' rule as

$$\eta = \Pr\left(G \mid \text{FLAG} = 1\right) = \frac{qa}{qa + \left(1 - q\right)b} \tag{S2.9}$$

With sufficiently accurate QAE/QPE classification at a fixed threshold, $p$ approaches the reference qualifying fraction $q$, and $\eta$ approaches one. At finite evaluation resolution, $p$ remains the operational probability of FLAG acceptance before QSEL, while $\eta$ gives the benchmark-level fraction of selected outcomes that meet the threshold. The value of $\eta$ does not determine whether a particular retrieved index satisfies the threshold. Ideal QSEL changes FLAG acceptance and the shots needed for detection but preserves $\eta$.

## Supplementary Note S3. Search-space and evaluation-register settings

Table S1 lists the selector-register requirements for the eleven nested ligand libraries. The occupancy selector grows with the library size, while the receptor-state, rotation, and translation selectors have fixed widths.

**Table S1.** Selector-register requirements across the eleven nested ligand libraries.

| $N_{OCC}$ | $N_{search}$ | Receptor | OCC | ROT | TRANS | Total |
|---|---|---|---|---|---|---|
| 1 | 32,768 | 1 | 0 | 5 | 9 | 15 |
| 2 | 65,536 | 1 | 1 | 5 | 9 | 16 |
| 4 | 131,072 | 1 | 2 | 5 | 9 | 17 |
| 8 | 262,144 | 1 | 3 | 5 | 9 | 18 |
| 16 | 524,288 | 1 | 4 | 5 | 9 | 19 |
| 32 | 1,048,576 | 1 | 5 | 5 | 9 | 20 |
| 64 | 2,097,152 | 1 | 6 | 5 | 9 | 21 |
| 128 | 4,194,304 | 1 | 7 | 5 | 9 | 22 |
| 256 | 8,388,608 | 1 | 8 | 5 | 9 | 23 |
| 512 | 16,777,216 | 1 | 9 | 5 | 9 | 24 |
| 1,024 | 33,554,432 | 1 | 10 | 5 | 9 | 25 |

Note. The columns Receptor, OCC, ROT, and TRANS give receptor-state, occupancy, rotation, and translation selector qubit counts, respectively. Total excludes the encoded data, evaluation, FLAG, and comparator-work registers. $N_{search}$ includes two receptor states, 32 rotation states, and 512 translations per occupancy.

Table S2 lists the amplitude thresholds and evaluation-register settings. For each base setting $m_0$, increasing the resolution multiplier $k$ changes the evaluation-register size $m$ while preserving the screening threshold $A_{0,\mathrm{thr}}$.

**Table S2.** Threshold settings and evaluation-register sizes for $k = 1, 2, 4$, and 8.

| $m_0$ | $A_{0,\mathrm{thr}}$ | $m$ ($k = 1$) | $m$ ($k = 2$) | $m$ ($k = 4$) | $m$ ($k = 8$) |
|---|---|---|---|---|---|
| 10 | 3.068e-03 | 10 | 11 | 12 | 13 |
| 11 | 1.534e-03 | 11 | 12 | 13 | 14 |
| 12 | 7.670e-04 | 12 | 13 | 14 | 15 |
| 13 | 3.835e-04 | 13 | 14 | 15 | 16 |
| 14 | 1.917e-04 | 14 | 15 | 16 | 17 |
| 15 | 9.587e-05 | 15 | 16 | 17 | 18 |
| 16 | 4.794e-05 | 16 | 17 | 18 | 19 |
| 17 | 2.397e-05 | 17 | 18 | 19 | 20 |
| 18 | 1.198e-05 | 18 | 19 | 20 | 21 |
| 19 | 5.992e-06 | 19 | 20 | 21 | 22 |
| 20 | 2.996e-06 | 20 | 21 | 22 | 23 |
| 21 | 1.498e-06 | 21 | 22 | 23 | 24 |
| 22 | 7.490e-07 | 22 | 23 | 24 | 25 |

Note. In the numerical benchmark, $A_{0,\mathrm{thr}} = \sin(\pi/2^{m_0})$ and $m = m_0 + \log_2(k)$.

**Supplementary Note S4. Gate-level and full-scale pose-overlap validation**

To assess whether the logical architecture in Figure 1 can be assembled and executed as a unitary circuit, a reduced 23-qubit prototype was constructed with Qiskit and evaluated using gate-level statevector simulation [4]. The prototype contains a six-qubit DATA register, one AUX qubit, eight configuration-selector qubits, four evaluation qubits, one FLAG qubit, and three reusable WORK qubits. The 256 selector states cover the indexed map, occupancy, rotation, and translation settings of the reduced model. Figure S1 presents the corresponding circuit schematic.

The circuit implements selector-controlled occupancy preparation, reversible rotation and translation, receptor-map encoding, QAE/QPE, a reversible comparator, and one QSEL iteration. For the four-qubit QPE evaluation register, the comparator threshold was set to 0.2. Both eigenphase acceptance regions were included, accepting the bins $y = 0$, 1, and 15. QSEL was constructed from the complete coherent state preparation and its inverse. The simulation output was compared with independently calculated reference amplitudes and probabilities.

The maximum absolute differences for the component and end-to-end comparisons are listed in Table S3. These results demonstrate functional agreement for a reduced gate-level implementation.

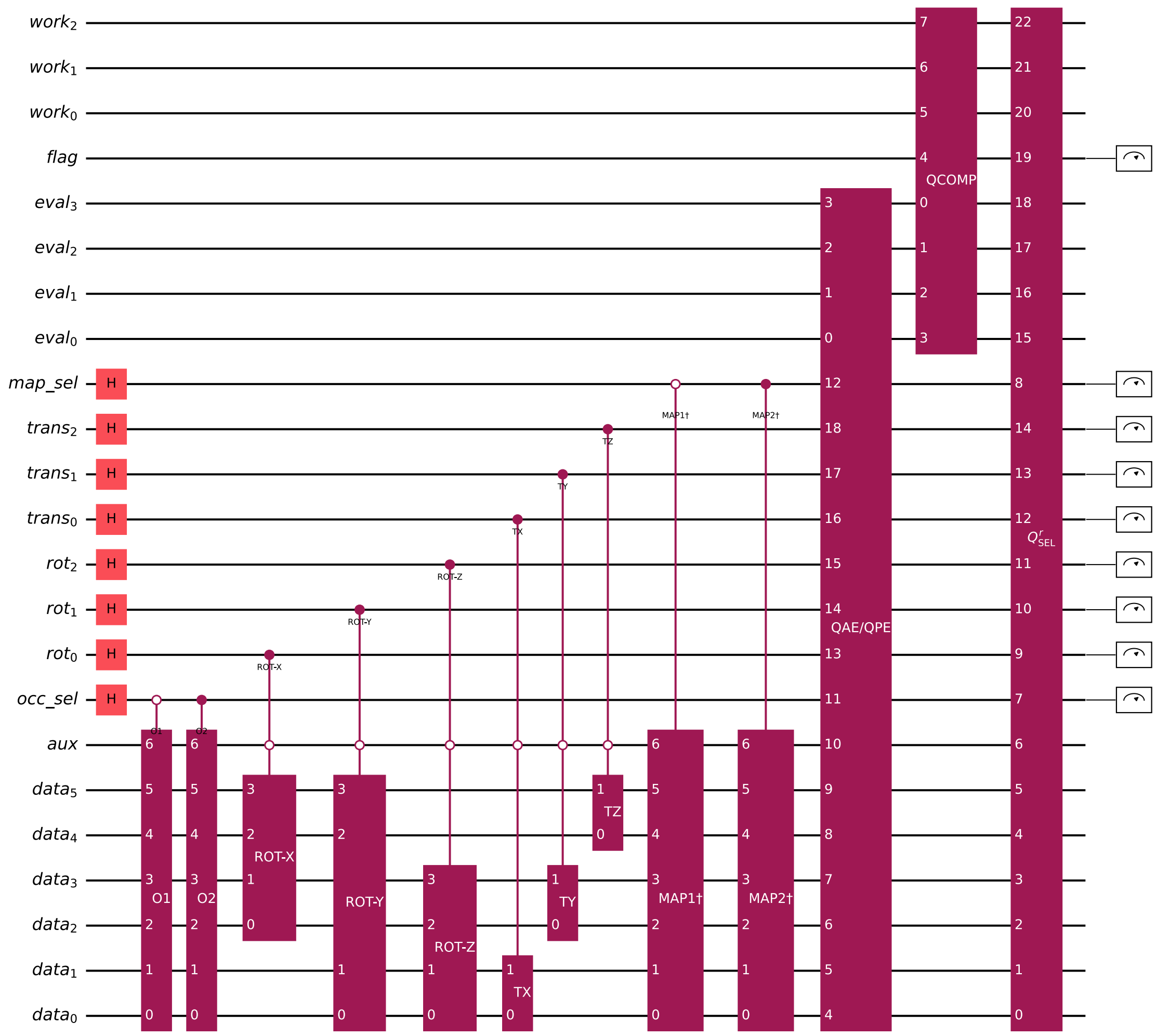

**Figure S1. Schematic of the 23-qubit circuit used for gate-level validation.** The reduced circuit includes eight selector qubits, six DATA qubits, one AUX qubit, four QPE evaluation qubits, one FLAG qubit, and three WORK qubits. SELECT, ROT, TRANS, QAE/QPE, QCOMP, and QSEL are shown as logical blocks without decomposition into elementary gates. FLAG and the selector registers are measured after QSEL.

**Table S3.** Agreement between independent analytical calculations and the 23-qubit gate-level statevector simulation.

| Circuit component | Maximum absolute error |
|---|---|
| Rotation and translation | $9.334 \times 10^{-14}$ |
| Inner-product amplitude $A_0$ | $1.128 \times 10^{-12}$ |
| QAE/QPE phase estimation | $5.651 \times 10^{-12}$ |
| QCOMP/FLAG acceptance | $2.198 \times 10^{-13}$ |
| Selector distribution before QSEL | $2.938 \times 10^{-14}$ |
| Selector distribution after QSEL | $3.454 \times 10^{-13}$ |
| FLAG probability before QSEL | $1.716 \times 10^{-12}$ |
| FLAG probability after QSEL | $2.136 \times 10^{-11}$ |

Note. All comparisons satisfied a numerical tolerance of $10^{-8}$.

**Full-resolution ROT/TRANS gate-level validation.** To validate at the quantum-gate level the ROT/TRANS coordinate transformations used in the present QAE implementation, we constructed a 29-qubit Qiskit Aer matrix-product-state simulation of the 32 × 32 × 32 grid. The circuit used 15 coordinate qubits, five ROT-selector qubits (32 selector states, 24 unique orientations), and nine TRANS-selector qubits (512 states). Its selector-bit assignments and transformation order matched those of the original QAE implementation. ROT was implemented using controlled swaps and bit flips. TRANS used five-qubit quantum Fourier transforms, controlled phase shifts, and inverse transforms to perform cyclic modulo-32 translations. For ROT, selector bits 4 to 0 controlled, in order, the 90° and 180° $z$-axis rotations, the 90° and 180° $y$-axis rotations, and the 90° $x$-axis rotation. For TRANS, the lowest three selector bits encoded $x$, the next three encoded $y$, and the highest three encoded $z$. On each axis, the least-significant bit contributed +1, the middle bit contributed +2, and a zero in the most-significant bit contributed −4 grid cells. The resulting shift was applied modulo 32.

To verify the Qiskit gate-level implementation, its output coordinates were compared with independently computed NumPy coordinate transformations for all 32 × 512 selector combinations at four initial coordinates, (3, 11, 25), (0, 0, 0), (31, 31, 31), and (0, 17, 31). All 65,536 basis-state checks passed, as did 256 one-axis TRANS checks. A 16-branch coherent interference test using an independently synthesized reversible reference correction yielded a return probability of 1.000000000000 and leakage of $8.154 \times 10^{-21}$. These tests verified the Qiskit ROT/TRANS circuit against an independent coordinate-based reference, rather than by directly running the original QAE program. Separately, the original QAE ROT/TRANS and FFT-based overlap calculation were validated against an independent non-FFT implementation, as detailed in the full-scale pre-QAE overlap validation below.

**Full-scale pre-QAE overlap validation.** Two implementations used the same normalized 19-qubit receptor MAP and ligand OCC vectors stored in .bin files. The production path used the original selector mapping, grid rotations, and FFT-based cyclic correlations. The independent non-FFT path applied explicit coordinate permutations for the 32 ROT states and modulo-32 coordinate shifts for the 512 TRANS states. It summed receptor-map values weighted by nonzero occupancy amplitudes over all eight channels. The unchanged auxiliary inner product was included in $A_0$.

All eight receptor systems, two states per receptor, 1,024 occupancies, 32 ROT states, and 512 TRANS states yielded 268,435,456 matched configurations. Both the encoded $A_0$ and the physical grid overlap were checked with absolute and relative tolerances of $10^{-9}$. No comparison failed. The maximum absolute difference in $A_0$ was $2.43 \times 10^{-17}$, and the overall root mean square error was $2.22 \times 10^{-18}$.

**Supplementary Note S5. Qualifying fraction in the constructed libraries**

Figure S2 shows the reference qualifying fraction $q$ calculated directly from grid-interaction scores at thirteen $A_{0,\mathrm{thr}}$ values for $N_{\mathrm{OCC}} = 1$, 16, 128, and 1024. The blue curves show medians across eight receptor systems, and shading indicates interquartile ranges (IQRs). Unlike the finite-resolution FLAG acceptance probability $p$ in main Figure 2, $q$ does not depend on $k$ or QSEL. This reference characterizes the score distribution of the constructed libraries. The values of $p$ and $\eta$ describe the operational selection behavior and its reliability.

At $A_{0,\mathrm{thr}} \approx 3.07 \times 10^{-3}$, the median $q$ was 0.983, 0.296, 0.181, and 0.0823 for $N_{\mathrm{OCC}} =$ 1, 16, 128, and 1024, respectively. Thus, library composition influences the fraction of configurations qualifying at the same threshold.

At $A_{0,\mathrm{thr}} \approx 7.49 \times 10^{-7}$, the corresponding median $q$ values were $5.95 \times 10^{-4}$, $3.72 \times 10^{-5}$, $4.65 \times 10^{-6}$, and $5.81 \times 10^{-7}$. Expressed as configuration counts, the median number of threshold-qualified configurations across the eight receptor systems was approximately 19.5 at each library size, consistent with near-constant-count behavior.

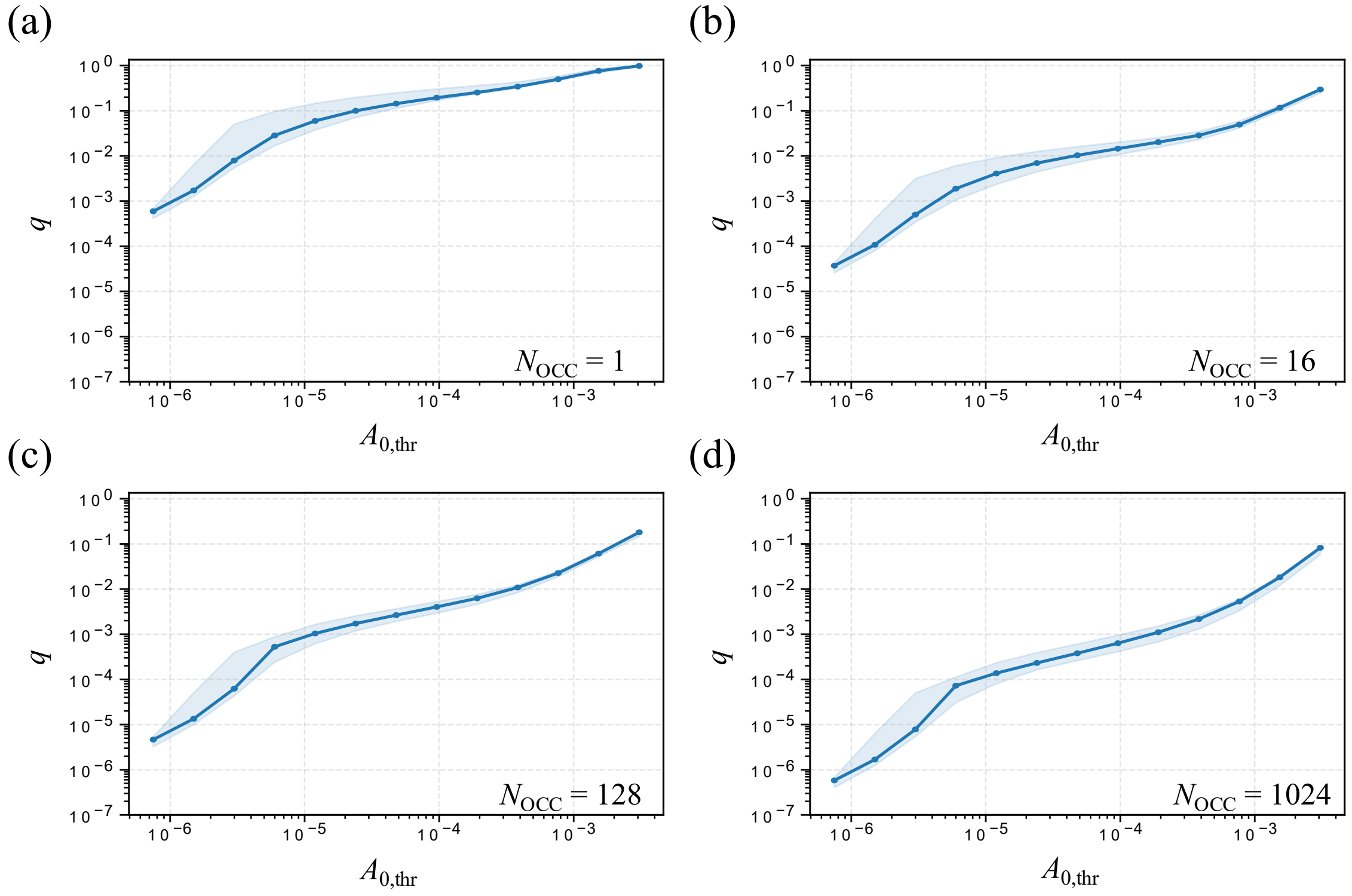

**Figure S2. Reference qualifying fraction** $q$ versus $A_{0,\mathrm{thr}}$ in the constructed nested libraries. Panels (a)–(d) correspond to $N_{\mathrm{OCC}} = 1$, 16, 128, and 1024, respectively. The blue solid curve shows the eight-receptor median, and the shaded band indicates the 25th–75th percentiles. The fraction $q$ is independent of $k$.

**Supplementary Note S6. FLAG-detection shots and logical costs across library sizes**

Figures S3 and S4 compare the same four $N_{\mathrm{OCC}}$ values at $m_0 = 14$ ($A_{0,\mathrm{thr}} \approx 1.92 \times 10^{-4}$), with $k = 1$, 2, 4, and 8. Figure S3 shows $S_{95}(r)$, the minimum number of shots needed to observe FLAG = 1 with at least 95% probability. In each panel, the QSEL iteration count $r$ starts at zero, allowing direct comparison with execution without QSEL. The maximum sampled iteration count could differ among resolution multipliers, and the curves terminate at their sampled limits.

Figure S4 reports $C_{95}(r)$ for these settings. More QSEL iterations can reduce $S_{95}(r)$, but each shot then requires $2r + 1$ QAE/QPE evaluation blocks. Consequently, the fewest shots and the lowest $C_{95}(r)$ need not occur at the same $r$.

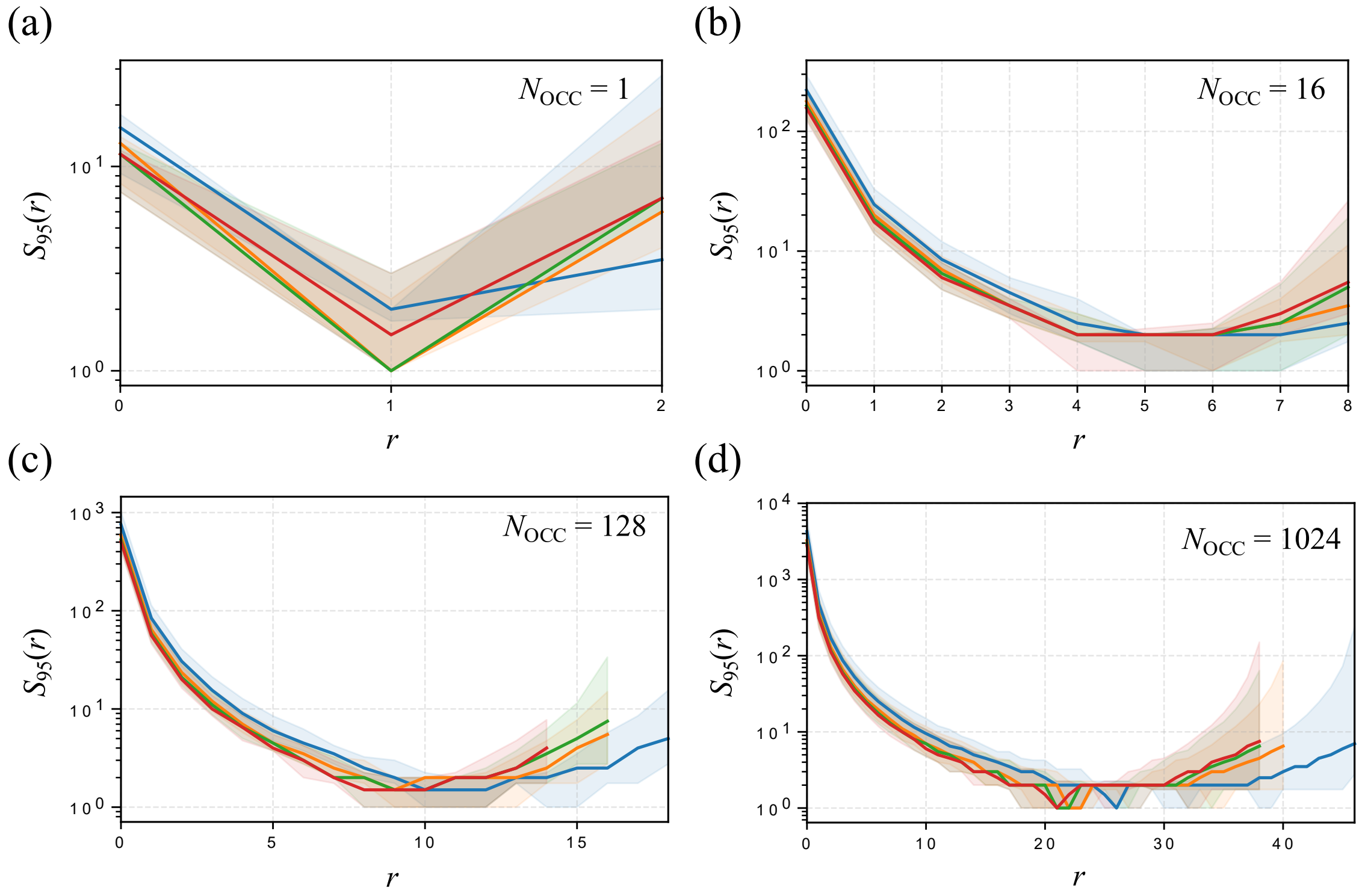

**Figure S3. Minimum shot count** $S_{95}(r)$ versus QSEL iterations $r$ for 95% FLAG detection. Panels (a)–(d) show $N_{\mathrm{OCC}} = 1$, 16, 128, and 1024 at $m_0 = 14$ ($A_{0,\mathrm{thr}} \approx 1.92 \times 10^{-4}$). Blue, orange, green, and red solid curves represent $k = 1$, 2, 4, and 8, respectively. Curves show medians across eight receptors. Shaded bands indicate the 25th–75th percentiles.

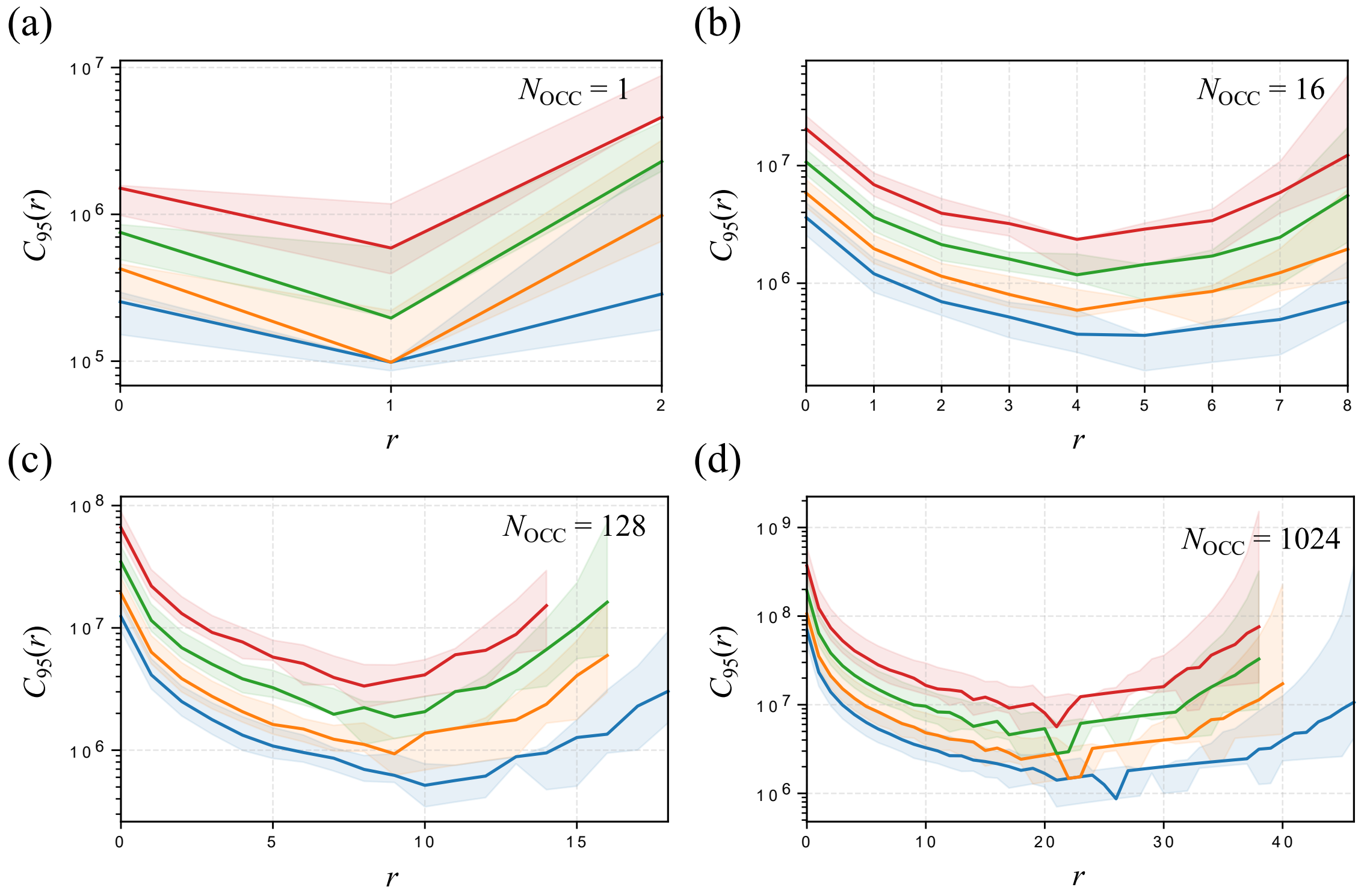

**Figure S4.** $C_{95}(r)$ versus QSEL iterations $r$ for FLAG detection at 95% probability. Panels (a)–(d) correspond to $N_{\mathrm{OCC}}$ = 1, 16, 128, and 1024 at $m_0$ = 14 ($A_{0,\mathrm{thr}} \approx 1.92 \times 10^{-4}$). Blue, orange, green, and red solid curves represent $k$ = 1, 2, 4, and 8, respectively. Curves show medians across eight receptors. Shaded bands indicate the 25th–75th percentiles. Costs are logical $Q$-operator calls.

### Supplementary Note S7. Minimum FLAG-detection costs across library sizes

Figure S5 shows $\min_r C_{95}(r)$ over the sampled nonnegative QSEL iteration counts as a function of $N_{\mathrm{OCC}}$. Together with main Figures 2 and 3, these results help assess the library-size dependence of QSEL costs at a chosen threshold and evaluation resolution.

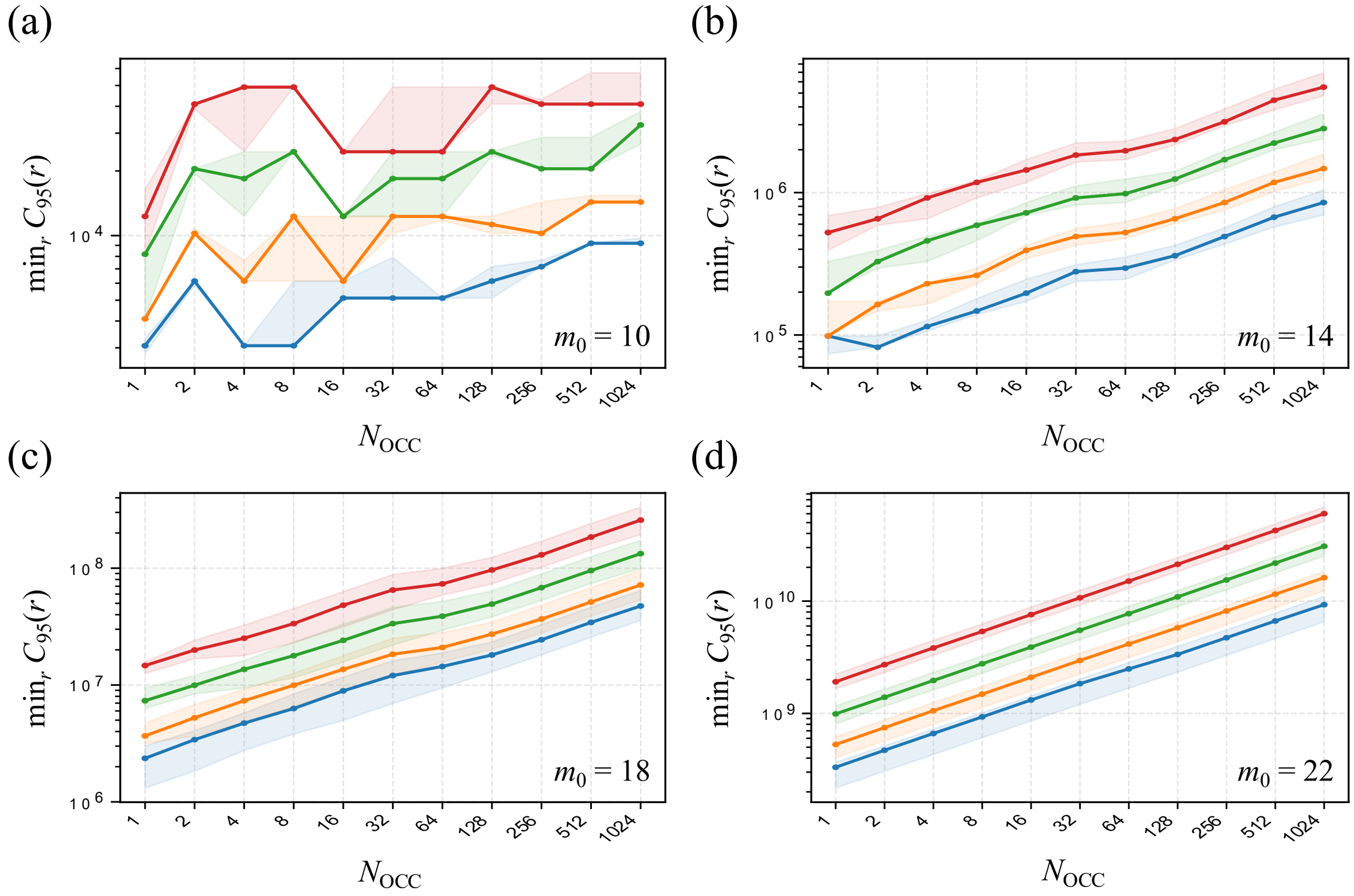

**Figure S5.** $\min_r C_{95}(r)$ versus $N_{\mathrm{OCC}}$ for FLAG detection at 95% probability. Panels (a)–(d) show $m_0$ = 10, 14, 18, and 22, respectively. For each receptor, $C_{95}(r)$ is minimized over the sampled $r$ values before calculating the median and IQR across eight receptors. Blue, orange, green, and red solid curves represent $k$ = 1, 2, 4, and 8, respectively. Shaded bands indicate the 25th–75th percentiles.